\documentclass{aa}  
\usepackage{graphicx}

\usepackage{txfonts}
\usepackage{xcolor}
\usepackage{float}
\usepackage{subcaption}
\usepackage{hyperref}
\usepackage{footnote}
\usepackage{soul}

\newcommand\fref[1]{(Fig. \ref{#1})}
\newcommand\tref[1]{(Tab. \ref{#1})}
\newcommand\sref[1]{(Sect. \ref{#1})}
\newcommand{\bluecite}[1]{{\color{blue}\cite{#1}}}
\newcommand{\bluecitep}[1]{{\color{blue}\citep{#1}}}
\newcommand{\Tufit}{\text{6723 }}
\newcommand{\Tufite}{\text{196 }}
\newcommand{\Tdfit}{\text{3394 }}
\newcommand{\Tdfite}{\text{264 }}
\newcommand{\Msol}{M$_\odot$ }

\newcommand{\BP}{\mathrm{B_P}}
\newcommand{\RP}{\mathrm{R_P}}
\newcommand{\V}{\mathrm{V}}
\newcommand{\change}{black}

\usepackage[normalem]{ulem}

\begin{document} 
    
   \title{Characterising the post-red supergiant binary system AFGL 4106 and its complex nebula with SPHERE/VLT}

   \author{G. Tomassini\inst{1}\thanks{\email{gabriel.tomassini@oca.eu}}
           \and E. Lagadec\inst{1}
           \and I. El Mellah\inst{4,5}
           \and R. D. Oudmaijer\inst{2,3}
           \and A. Chiavassa\inst{1}
           \and M. N'Diaye\inst{1}
           \and P. de Laverny\inst{1}
           \and N. Nardetto\inst{1}
           \and A. Matter\inst{1}
          }

   \institute{Université Côte d'Azur, Observatoire de la Côte d'Azur, CNRS, Laboratoire Lagrange, France
    \and 
    Royal Observatory of Belgium, Ringlaan 3, 1180 Brussels, Belgium
    \and
    School of Physics \& Astronomy, University of Leeds, Woodhouse Lane, LS2 9JT, Leeds, UK
    \and
    Departamento de Física, Universidad de Santiago de Chile, Av. Victor Jara 3659, Santiago, Chile
    \and
    Center for Interdisciplinary Research in Astrophysics and Space Exploration (CIRAS), USACH, Chile
             }

   \date{}

  \abstract
    {Mass loss in evolved massive stars plays a critical role in shaping their circumstellar environments and enriching the interstellar medium. In binary systems, stellar interactions can further complicate this process, affecting stellar evolution, stellar yields, and nebular morphology.}
   {We aim to characterise the physical and morphological properties of the binary system AFGL 4106, which is composed of two evolved massive stars. Understanding its mass-loss processes and circumstellar environment offers insight into the late stages of stellar evolution in massive binary systems.}
 {We obtained high-angular-resolution, high-contrast imaging using VLT/SPHERE with ZIMPOL (optical) and IRDIS (near-infrared) across multiple filters. We used aperture photometry to extract the spectral energy distributions (SEDs) of each star and applied radiative-transfer modelling to study the system and its surrounding dusty environment. }
   {The observations resolve both components of the binary and unveil a complex, dusty nebula featuring asymmetric structures and cavities. SED fitting yields stellar temperatures of T$_1$ = \Tufit $\pm$ \Tufite K and T$_2$ = \Tdfit $\pm$ \Tdfite K, along with bolometric luminosities of L$_1 = (7.9 \pm 0.18) \times 10^4$ L$_\odot$ and L$_2 = (3.8 \pm 0.11) \times 10^4$ L$_\odot$. These values support the classification of the primary as being in a post-red-supergiant (post-RSG) phase and the secondary as an active red supergiant (RSG). The luminosity ratio, combined with the inferred radii, indicates that both stars are at close yet distinct stages of their evolution. The binary is surrounded by an extended shell whose asymmetric morphology and large-scale features suggest interaction with the stellar winds and interstellar medium (ISM), and possibly the presence of a third, undetected companion.}
   {These observations provide the first resolved view of AFGL 4106's system and its dusty envelope. Our analysis sets constraints on the physical properties and evolutionary status of the system. This work contributes to our understanding of mass-loss processes in massive binaries and the shaping of nebulae around evolved stars.}

    \titlerunning{AFGL 4106: a Post-RSG binary system seen with SPHERE/VLT}
    \authorrunning{G. Tomassini et al.\fnmsep\ }
   \maketitle

\section{Introduction}
Mass loss from massive stars governs the chemical enrichment of the interstellar medium. Post-red supergiants (post-RSGs) are prime laboratories in which to study the morphology, composition, and kinematics of recently ejected material, and to reconstruct mass-loss histories, since they can undergo very strong outflows (up to $\sim10^{-4}\,M_{\odot}\,\mathrm{yr}^{-1}$; \bluecite{Mauron2011}). The class is small: only a few well-studied examples exist, such as IRC~+10420 \bluecitep{Koumpia2022} and the 'Fried Egg' nebula \bluecitep{Lagadec2011,Koumpia2020}. For a recent overview of post-RSGs and their circumstellar environments, see \bluecite{Gordon2019}; \textcolor{\change}{and for RSGs, their mass loss, and subsequent evolution, see \bluecite{VanLoon2025}.}

 The complex nebula shapes observed around such objects can be caused by several factors, such as magnetic fields or rotation, but it is often thought that the presence of one or more companions is the main source of the angular momentum excess necessary for their shaping \bluecitep{Balick2002}. 
 Observational studies and models have shown that a large proportion of stars --especially solar-type and massive stars-- reside in binary or multiple systems \bluecitep{Whitworth2015, Fuhrmann2017, Merle2024}. Companions must have an impact on the mass-loss compositions and rate from massive stars. Studying the mass loss history of post-RSGs and the influence of binaries is key to understanding matter ejection from massive stars before the supernova explosion.

AFGL 4106 provides an excellent case study for investigating how binarity influences mass loss in massive stars. It was examined in the late 1990s by \bluecite{Molster1999} and \bluecite{vanLoon1999}, both of which reached similar conclusions: the system is a binary, consisting of an evolved star --likely a post-RSG-- and a less evolved companion, embedded within a circumbinary nebula produced by one or both stars. The primary (star 1) is estimated to have a temperature of 7250 $\pm$ 250 K, while the companion (star 2) has a temperature of 3750 $\pm$ 250 K. The binary system was first resolved in 2006 using Hubble Space Telescope (HST) observations, leading to a separation measurement of 0.3" with a position angle (P.A.) of 270° \bluecitep{Bobrowsky2006}. 

The Spectro-Polarimetric High-contrast Exoplanet REsearch (SPHERE) on the Very Large Telescope provides an exquisite $ \sim$ 20 milliarcsec angular resolution in the optical and $\sim$ 50 milliarcsec in the near-infrared, enabling us to better characterise the central stars of the system. SPHERE's coronagraph and polarimetric capabilities enable a deep and detailed study of the nebula around the central binary system, allowing, for the first time, a study of the mass-loss history  of a resolved binary system including a post-RSG star.

In this paper, we present a detailed analysis of the evolved massive binary system AFGL 4106; it is based on high-angular-resolution, high-contrast observations obtained with VLT/SPHERE in both direct and polarimetric imaging modes across multiple optical and near-infrared wavelengths. These data reveal the morphology and scattering properties of the surrounding dusty nebula, putting constraints on the mass-loss history of the system.

This paper is structured as follows. In Section~\ref{sec:Observations}, we describe the SPHERE observations and data reduction. Section~\ref{sec:Results} presents the morphological and photometric analysis, including the spectral energy distributions of each component; and in it we discuss the radiative-transfer modelling used to constrain the stellar and dust parameters. In Section~\ref{sec:Discussion}, we interpret the results in the context of stellar evolution, mass-loss history, and nebular shaping mechanisms. Finally, our conclusions are summarised in Section~\ref{sec:Conclusion}.

\section{Observations and data reductions}
\label{sec:Observations}
\subsection{SPHERE observations}
Observations were carried out with the extreme \textcolor{\change}{adaptive} optics imager and polarimeter SPHERE \bluecitep{beuzit2019} installed at the Unit Telescope 3 \textcolor{\change}{Nasmyth} focus of the Very Large Telescope (VLT). Its adaptive optics module SAXO \bluecitep{fusco2006} provides a correction from the atmosphere turbulence to reach the highest angular resolution possible. The main \textcolor{\change}{component} of the adaptive optics system is the Shack-Hartmann wave-front sensor (WFS), with 41$\times$41 actuators working at a frequency of 1380 Hz on the deformable mirror. 

We obtained intensity images of AFGL\,4106 using the SPHERE instruments ZIMPOL (optical, 0.5-0.9 $\mu$m) and IRDIS (near-infrared, 0.95-2.32 $\mu$m). To enhance contrast in the ZIMPOL data, we acquired polarimetric maps, as dust-scattered light is polarised, while stellar emission is not, revealing the circumstellar environment at flux levels several orders of magnitude fainter than the stars. For selected optical filters, we combined polarimetry with a 77.5 mas coronagraph to further suppress the primary star's intensity and improve contrast. Additional data were obtained without a coronagraph to probe the innermost regions of the system. IRDIS observations were performed in classical imaging mode, since no coronagraph was large enough to mask both stars\textcolor{\change}{;} we increased the exposure time and allowed the central region to saturate in some IRDIS frames to better detect the nebula. A summary of the SPHERE observations is provided in Tab.~\ref{tab:log}.

Both ZIMPOL and IRDIS data were reduced using the SPHERE data-reduction centre \bluecitep{delorme2017}, which makes use of the Zurich ZIMPOL pipeline \bluecitep{Schmid2018} and the SPHERE pipeline \bluecitep{Pavlov2008} for the basic reduction steps. The SPHERE data centre makes use of custom-designed python routines to provide the ZIMPOL polarisation map. In \textcolor{\change}{polarisation} mode, ZIMPOL measures the Stokes parameters Q and U; the \textcolor{\change}{polarisation} maps are then computed using the following equations: 

\begin{equation}
    \qquad p_L = \sqrt{q^2 + u^2}; \quad I = \frac{I_Q+I_U}{2}; \quad \theta = \textcolor{\change}{\arctan\left(\frac{q}{u}\right)}
    \label{eq:pol}
\end{equation}

\begin{equation}
     \text{with: }\hspace{1.5cm} q = \frac{Q_+ + Q_-}{2I_Q}; \quad u = \frac{U_+ + U_-}{2I_U}
,\end{equation}and where $p_L$ is the polarized intensity, $I$ the total intensity, $\theta$ the polarization angle, $q$ and $u$ the normalized Stokes parameters, $I_Q$ and $I_U$ the total intensity measured in the Q and U cycles, respectively, and $Q_+$, $Q_-$, $U_+$ and $U_-$  the Stokes parameters measured in each half cycle.

\subsection{\textcolor{\change}{Auxiliary} photometry}
The spectral energy distribution (SED) of the system is affected by the properties of the two \textcolor{\change}{stars} and the interaction of their light with dust around them.
To construct the SED of AFGL 4106, \textcolor{\change}{we} retrieved archival multi-wavelength photometric data covering both the optical and infrared ranges. This broad spectral coverage allows us to characterise the contributions from both the stellar components and the surrounding dust. Additionally, we incorporated ISO-SWS spectra \bluecitep{Sloan2003} to complement the infrared portion of the SED. The spectra were integrated over ISO filter response curves to extract corresponding photometric points.
The data we used are presented in Table \ref{table:Data_SED}.

\section{Analysis}
\label{sec:Results}
\subsection{The central stars}

\begin{figure}[h!]
    \centering
    \includegraphics[width=\linewidth]{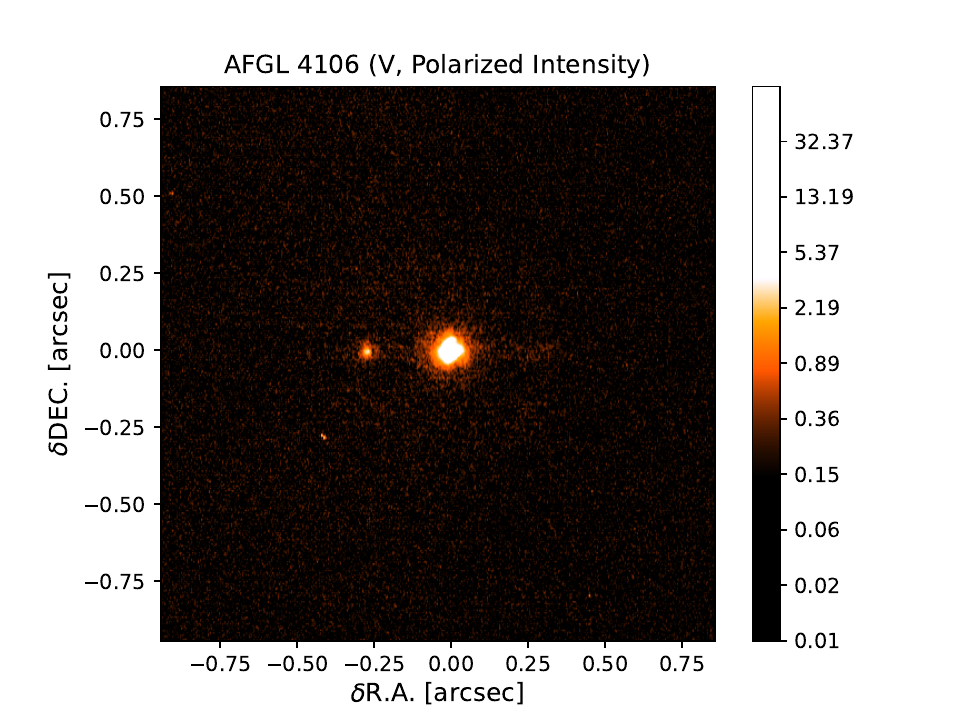}
    \caption{ZIMPOL polarised intensity map of AFGL 4106 in filter V.}
    \label{fig:V}
\end{figure}

From Hubble Space Telescope (HST) imaging, the separation between the two central stars has been estimated as 0.3" \bluecitep{Bobrowsky2006} with a position angle of $\sim$270$^\circ$ for the secondary. However, the theoretical diffraction limit ($\lambda/\mathrm{D}$) of HST at 1.644 $\mu$m is 0.141". This implies that the separation of 0.3" is just above the Nyquist limit. With our SPHERE ZIMPOL images, we reached a theoretical diffraction limit of 0.014" in the V band \fref{fig:V}, which is about one order of magnitude better than HST resolution. This allows us to obtain a better estimation of the separation. We performed a photocentre fit on each components in the V-band image in order to measure the separation. For this, we used the iteratively-weighted centre-of-gravity (IWCoG) technique \bluecitep{Vyas2009}, where the photocentre is defined as

\begin{equation}
    \hspace{2cm}(\text{x}_\text{c},\text{y}_\text{c}) = \frac{\sum_{ij}\text{X}_{ij}I_{ij}W_{ij}}{\sum_{ij}I_{ij}W_{ij}} \quad 0<i,j<1023,
\end{equation}where (x$_\text{c}$, y$_\text{c}$) are the photocentre coordinates, X is the position of the pixel, I the value associated with the pixel, and W a Gaussian weight, calculated using a radial profile of the star. We finally find a separation of 0.272 $\pm$ 0.017 arcsec with a position angle of $\sim$90°. \textcolor{\change}{At a distance of 3.19$^{+0.45}_{-0.19}$ kpc (see Sect.~\ref{sec:Distance}), this corresponds to a physical separation of 865 $\pm$ 102 au.}  Our measurement differs from that reported by \bluecite{Bobrowsky2006}. This discrepancy arises from the choice of the reference star, which introduces a 180 degree° ambiguity in the position angle. They indeed used star 2 as a reference, while we used star 1 (the one in the east in all images).

\subsection{The nebula}
\label{sec:Nebula}

\begin{figure}[h!]
    \centering
    \includegraphics[width=\linewidth]{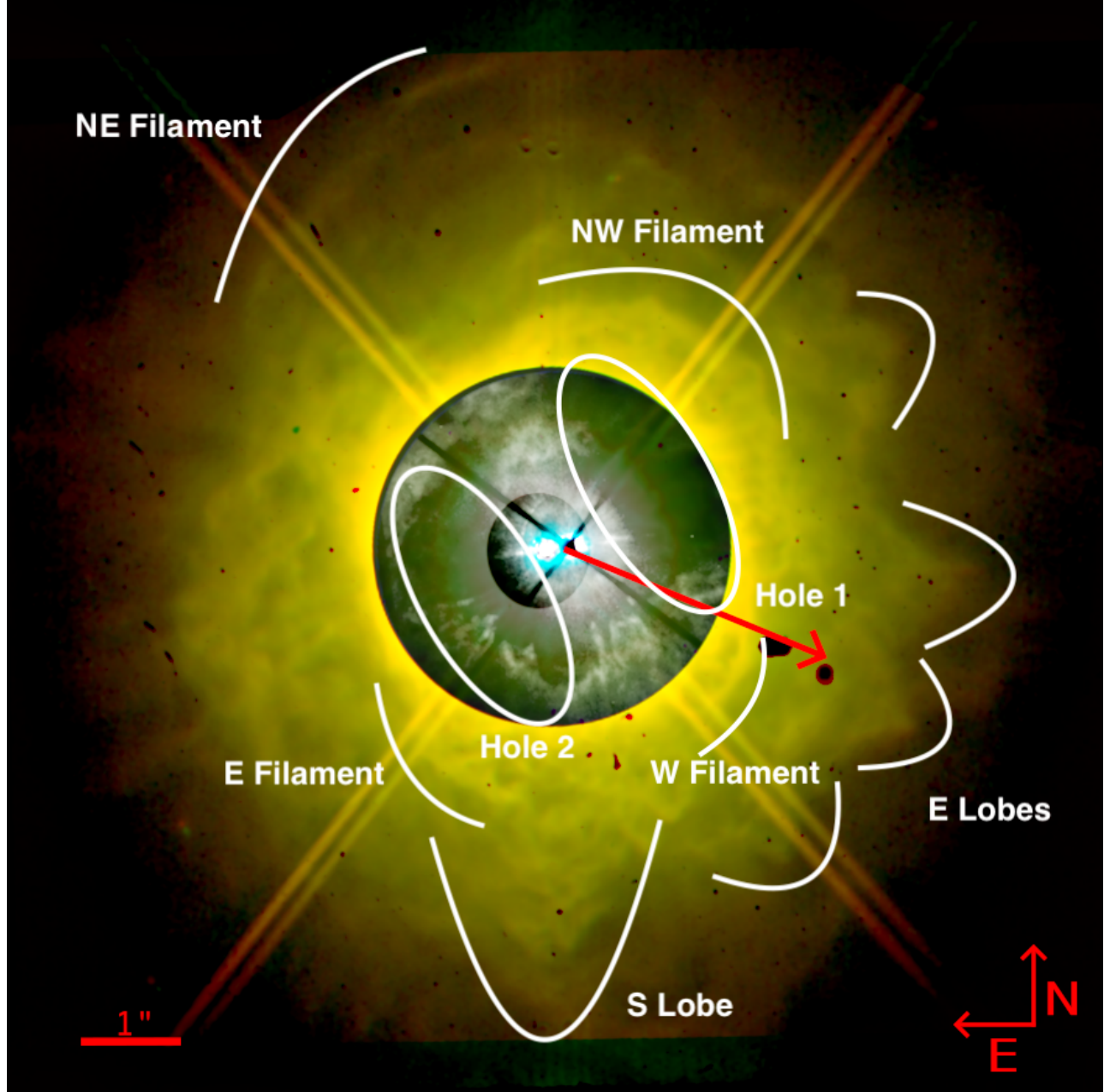}  
    \caption{Three-colour image of AFGL 4106 in Band I\_PRIM, J and H. ZIMPOL image is re-scaled to the IRDIS FoV. The colour and contrast are enhanced differently in the inner and outer parts of the nebula. The red arrow indicates the direction of the proper motion of the system.}
    \label{fig:3col}
\end{figure}
\subsubsection{ZIMPOL optical-imaging polarimetry}

\begin{figure}[h!]
    \centering
    \includegraphics[width=\linewidth]{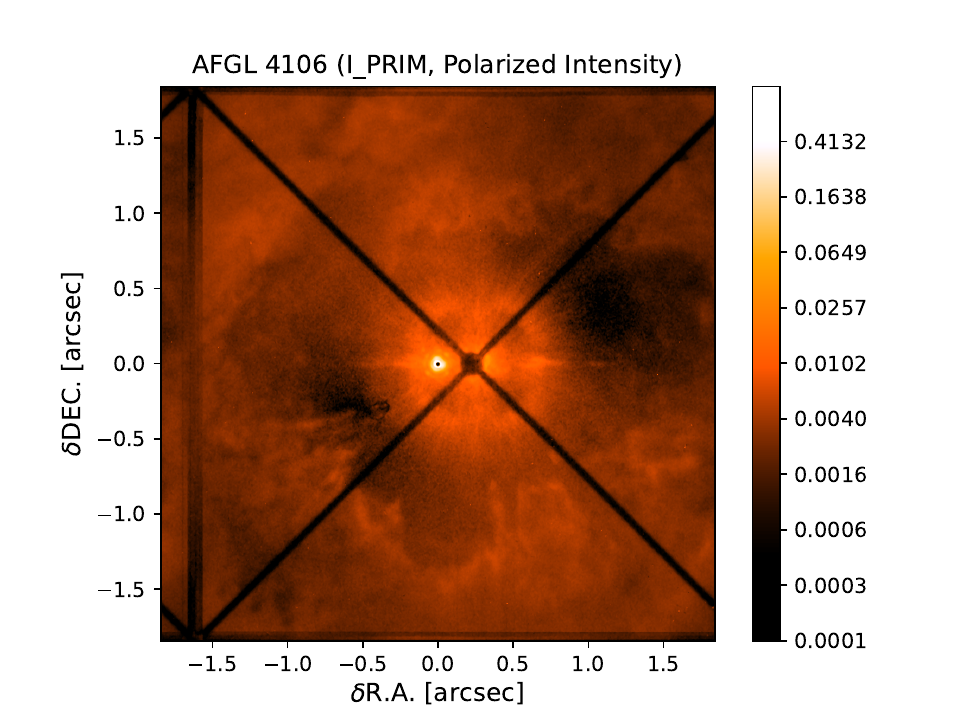} 
    \caption{ZIMPOL polarised-intensity map of AFGL 4106 in filter I\_PRIM.}
    \label{fig:I_PRIM}
\end{figure}

With its narrow field of view ($\sim3.5^{\prime\prime} \times 3.5^{\prime\prime}$) and high angular resolution (about 20 mas, \bluecite{Schmid2018}), observations with the ZIMPOL instrument reveal the morphology of the inner part of the nebula.  To enhance the contrast, we used a coronagraph (seen as a black circle with spiders) and polarimetric measurements. The light from the star is unpolarised, while dust scatters this light. Images in polarised light enabled us to partially mask the central star and reveal the dust around it. Figure~\ref{fig:I_PRIM} presents the polarised-intensity image obtained with the ZIMPOL I\_PRIM filter (central wavelength: 789.7 nm).

The central circular shape, with a diameter of $\sim$1 arcsec, is the adaptive optics correction halo and, therefore, not physical.
The ZIMPOL polarised-intensity image \fref{fig:I_PRIM} highlights two prominent cavities, which are referred to as holes 1 and 2. These cavities, each measuring approximately $2''$ in length and $1''$ in width, are positioned symmetrically on either side of a bright northeast-southwest axis.  

In the polarised-intensity images, the two stars appear embedded in a non-uniform cavity of $\sim$2 arcsec in diameter, including the two voids described before.  The inner edges of this cavity are clearly non-spherical, and a dichotomy is observed between the northwest rim, where four loops of $\sim$0.3 arcsec in diameter are observed, and the southeast displays five comet-like features pointing towards the centre.
The external parts of the observed nebula are more uniform, albeit still clumpy.

\subsubsection{IRDIS near-infrared imaging}
\label{sec:IRDIS}

The IRDIS intensity images reveal the outer parts of the nebula. As no coronagraphs were large enough to mask the central stars, the central stars were heavily saturated in some sequences to reveal the nebula.

\begin{figure}[h!]
    \centering
    \includegraphics[width=\linewidth]{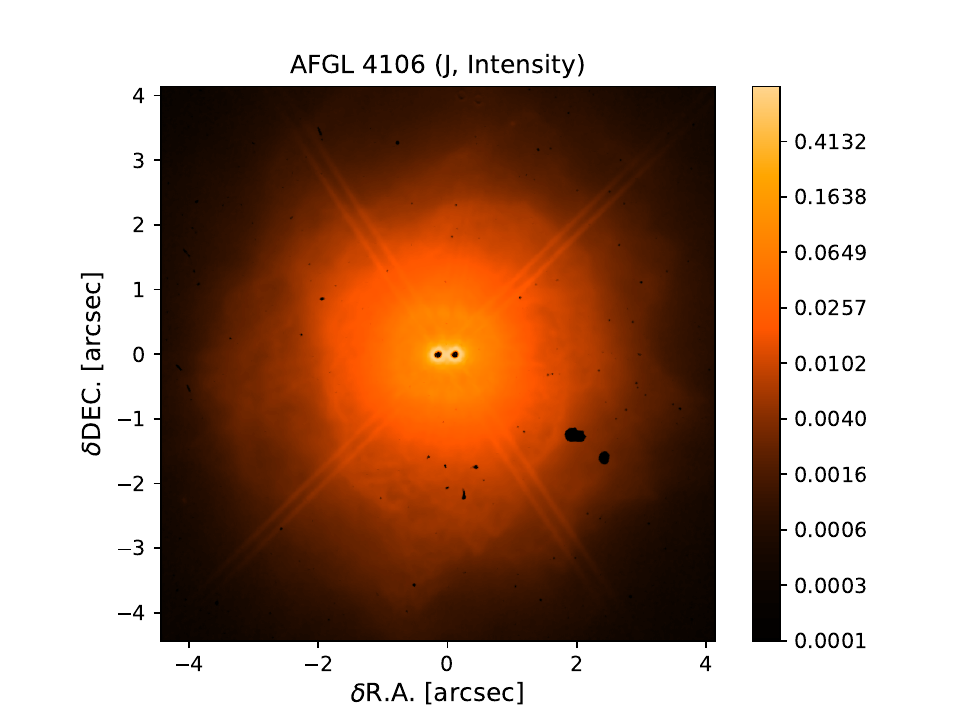}
    \caption{IRDIS intensity map of AFGL 4106 in filter J.}
    \label{fig:J}
\end{figure}

The  nebula has a diameter of approximately $8.6^{\prime\prime}$. It exhibits a variety of intricate structures at multiple scales.
Inside the nebula, the saturated stars appear has black dots. The two double lines crossing the stars represent the spider of M2. The $\sim$1.5 arcsec diameter circle is the AO correction halo.
Two filaments are observed on the southeast and west sides of the nebula, about 2 arcsec from the centre, clearly visible in \fref{fig:J}. Another one is present in the northeast, about 3 arcsec away from the centre. Finally one is present in the northwest about 2 arcsec away from the centre. The gaps between the filaments are not empty, but rather clumpy.

On a larger scale, in the IRDIS images we observe sharp, triangle-shaped lobes exiting the central star radially. They are observed at various distances between 1.5 and 2.0 arcsec from the central star. They appear to be sharper towards the south and west of the nebula.

\textcolor{\change}{They are similar to the lobes observed on CPD-568032 \bluecitep{Chesneau2006} or in the starfish nebula \bluecitep{Sahai2005} and are also similar to the sporadic, clumpy ejecta imaged around VY CMa \bluecitep{Humphreys2021}.} Mostly visible in filter J \fref{fig:J} and continuum J \fref{fig:CntJ}, these lobes are not visible at larger wavelengths. The most prominent structure is the southern one denoted as S lobe on Fig. \ref{fig:3col}, which is $\sim$2" long. The other lobes are smaller, and all are $\sim$0.5" long.

\textcolor{\change}{
Throughout the nebula, we observe filamentary structures that likely arise from interactions between the circumstellar material and the interstellar medium (ISM). \bluecite{vanLoon1999} reported similar filaments on larger scales and interpreted them as bow shocks produced by the motion of AFGL 4106 through the ISM. Using Gaia DR3 astrometry \bluecitep{Gaia2023}, we derive a proper motion of $\mu_{\alpha} = -6.68 \pm 0.33$~mas\,yr$^{-1}$ and $\mu_{\delta} = -1.46 \pm 0.35$~mas\,yr$^{-1}$, corresponding to a total motion of 6.84~mas\,yr$^{-1}$ with a position angle of $257^\circ$. This direction is indicated by the red arrow in Fig.~\ref{fig:3col}. However, the fact that the filaments are not spatially aligned with the proper-motion direction disfavours the bow-shock interpretation. Instead, these features are more likely the relics of past mass-ejection events from the system, subsequently shaped by the interaction with the surrounding ISM.
}

\begin{figure}[h!]
    \centering
    \subfloat[]{\label{fig:contour}\includegraphics[width=0.8\linewidth]{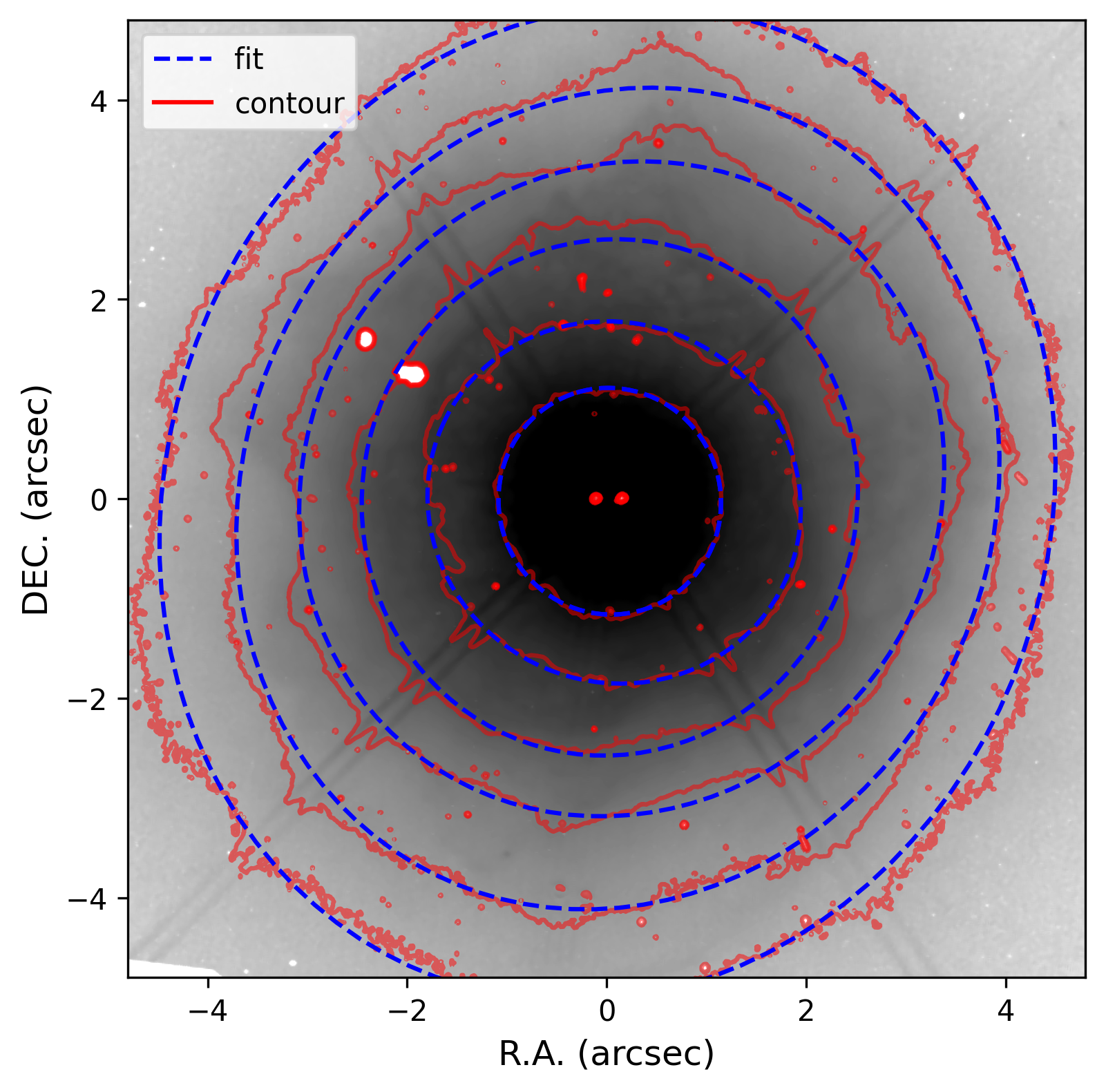}}\\
    \subfloat[]{\label{fig:ex}\includegraphics[width=0.8\linewidth]{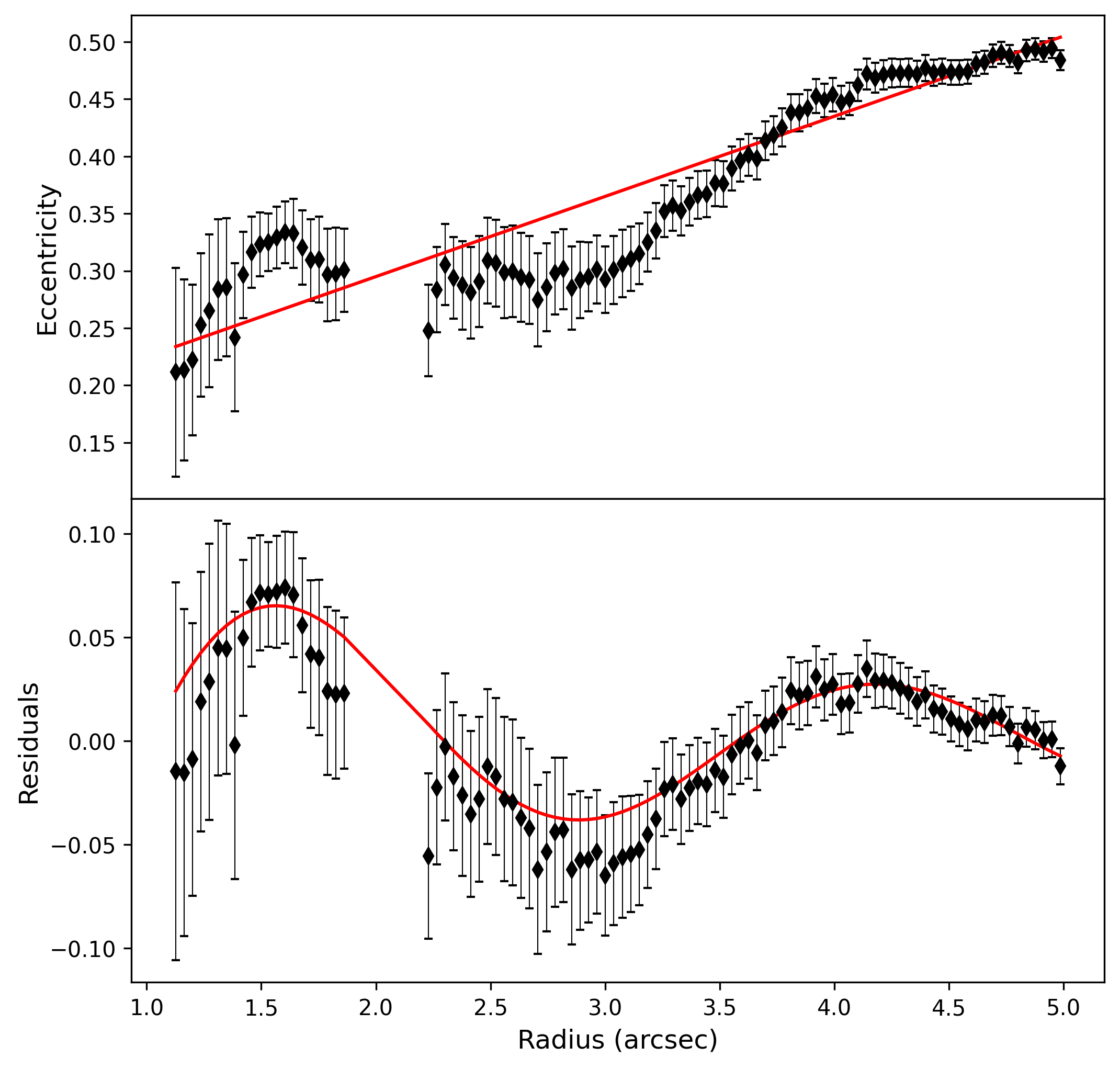}} 
    \caption{Zoomed-in view of intensity contours of AFGL 4106 nebula in J band is shown in Figure~\ref{fig:contour}, along with the corresponding eccentricity of these contours in Figure~\ref{fig:ex}. The top panels present the raw data, while the bottom panels show the residuals after subtracting a linear fit.}
    \label{fig:el}
\end{figure}

The iso-intensity map of the nebula displays variations in the shape of the \textcolor{\change}{ejecta} \fref{fig:contour}. Looking up at the eccentricity of these iso-intensity curve, $e=\sqrt{1-\frac{a^2}{b^2}}$,  where a is the semi-major axis and b the semi-minor axis, and $e$ seems to change with distance to the central star. We observe oscillations with a pseudo-period of $\sim$2.560 arcsec, given that we only have 1.5 periods. The gap observed between 1.8 and 2.2 arcseconds corresponds to a sharp variation of ellipticity, likely caused by a rotation of the elliptical contours by $\pi/2$, resulting in a swap between the semi-major and semi-minor axes. This suggest a change in the ejecta preferential direction across time.

Assuming a typical stellar-wind velocity of 20-30 km/s
\bluecitep{Molster1999} and a distance of 3.19 kpc (see discussion in Sect.~\ref{sec:Distance}), we estimate oscillation periods ranging from approximately 1300 to 1950 years, depending on the adopted wind speed.

\subsection{Spectral energy distribution of both stars}
Using an ISO-SWS spectrum covering the 2.36-45.3 $\mu$m range, \bluecite{Molster1999} previously estimated the temperatures of the system's components through SED fitting, finding a temperature of 7250$\pm$250 K for the primary and 3750$\pm$250 K for the secondary, concluding that the binary system consists of an  A-F  and \textcolor{\change}{an} M type star of similar luminosity. 

Using the first resolved observations of both components, we performed aperture photometry on our SPHERE data to isolate the SED of each star in the system across various optical and near-infrared wavelengths. Flux ratios were derived from the SPHERE data by integrating the flux within circular apertures of radius $1.22 \cdot \lambda / D$, corresponding to the diffraction-limited core and ensuring that the full dispersed flux is captured. Among the 16 available filters, only five provide a suitable combination of PSF quality and unsaturated stellar images, allowing us to perform aperture photometry on both stars. 

Since SPHERE data are not flux-calibrated, the total flux for each star was obtained from photometric \textcolor{\change}{auxiliary} data. These fluxes were corrected from the interstellar reddening using the PyAstronomy unred routine \bluecitep{PyA2019}, making use of the \bluecite{Fitzpatrick1999} parametrisation to estimate the wavelength dependence of interstellar extinction. This parametrisation uses spectral observation to derive the shape of the \textcolor{\change}{extinction} curve, fitted to the following \textcolor{\change}{formula:} 

\begin{equation}
\begin{aligned}
    \frac{\text{E}(\lambda - \text{V})}{\text{E}(\text{B} - \text{V})} &= \text{C}_2 + \text{C}_1 \lambda^{-1} 
    + \frac{\text{C}_3}{\left[ \lambda^{-1} - \frac{\lambda_0^{-2}}{\lambda^{-1}} \right]^2 + \gamma^2} \\
    &\quad + \text{C}_4 \left[ 0.539(\lambda^{-1} - 5.9)^2 + 0.0564(\lambda^{-1} - 5.9)^3 \right]
\end{aligned}
,\end{equation}

\textcolor{\change}{\\ where} E($\lambda$-V) is the extinction at wavelength $\lambda$, E(B-V) the extinction in the B and V bands, and

\begin{equation}
    \hspace{2.5cm} \left\{
    \begin{aligned}
        \text{C}_1 &=  -0.824 + 4.717 \times R^{-1} \\
        \text{C}_2 &= 2.030 + 3.007 \times \text{C}_1 \\
        \text{C}_3 &= 3.23 \\
        \text{C}_4 &= 0.41 \\
        \lambda_0^{-1} &= 4.595 \, \mu\text{m}^{-1}\\
        \gamma &= 1.051 \, \mu\text{m}^{-1}
    \end{aligned}
    \right.
\end{equation}

\noindent for the Milky Way.

The interstellar extinction value derived by \bluecite{Molster1999}, \( E(\mathrm{B} - \V) = 1.07 \) mag, was compared with the value derived by de Laverny (in preparation, private communication). This latter estimate was obtained using Gaia/GSP-Spec atmospheric parameters \bluecitep{Recio2023} to compare observed and theoretical stellar colours in the Gaia bands and thus determine the total reddening along the line of sight. For AFGL 4106, GSP-Spec reports \( T_\mathrm{eff} = 6019 \) K, \( \log g = 1.75 \), and \([{\rm M/H}] = -0.57\) dex, adopting the calibrations as a function of T$_\text{eff}$ described \textcolor{\change}{in} \bluecite{Recio2023,Recio2024} . With these parameters, the expected intrinsic colour is \((\BP - \RP)_\mathrm{0} = 0.65 \pm 0.013\) mag, while the observed Gaia/DR3 colour is \((\BP - \RP)_\mathrm{obs} = 2.46 \pm 0.01\) mag, yielding \( E(\BP - \RP) = 1.809 \pm 0.013 \) mag.  

It is worth noting that that GSP-Spec has analysed a Gaia/RVS spectrum that is actually a combined spectrum of the unresolved binary system, which may slightly affect the atmospheric parameter determination. Moreover, as AFGL 4106 is likely variable, the derived extinction is estimated from an average spectrum obtained over the $\sim$2.5 years of Gaia/DR3 observations. Using our adopted stellar temperature (\( T_\mathrm{eff} = 6723 \) K, see \sref{sec:Star_prop}), the expected intrinsic colour becomes \((\BP - \RP)_\mathrm{0} \simeq 0.5\) mag, leading to a slightly higher reddening, \( E(\BP - \RP) \approx 1.95 \). This difference remains small and does not significantly affect the other derived parameters, such as luminosity, but should be mentioned.  

To facilitate comparison with the extinction from \bluecite{Molster1999}, we converted \( E(\BP - \RP) \) to \( E(B - V) \) using the extinction coefficients \( A(\BP)/A(\V) \) and \( A(\RP)/A(\V) \) from the ESA/Gaia photometric system documentation \bluecitep{Riello2021}. Therefore, we have

\begin{equation}
\hspace{1cm}
\begin{aligned}
    E(\BP-\RP)  &= (\BP - \RP) - (\BP-\RP)_0 \\
                &= A(\BP) - A(\RP) \\
                &= \left( \frac{A(\BP)}{A(V)} - \frac{A(\RP)}{A(V)} \right) \times R_V \times E(B-V)\\
                &\simeq 1.392 \times E(B-V)
\end{aligned}
\end{equation}

\begin{figure}[ht]
    \centering
    \includegraphics[width=\linewidth]{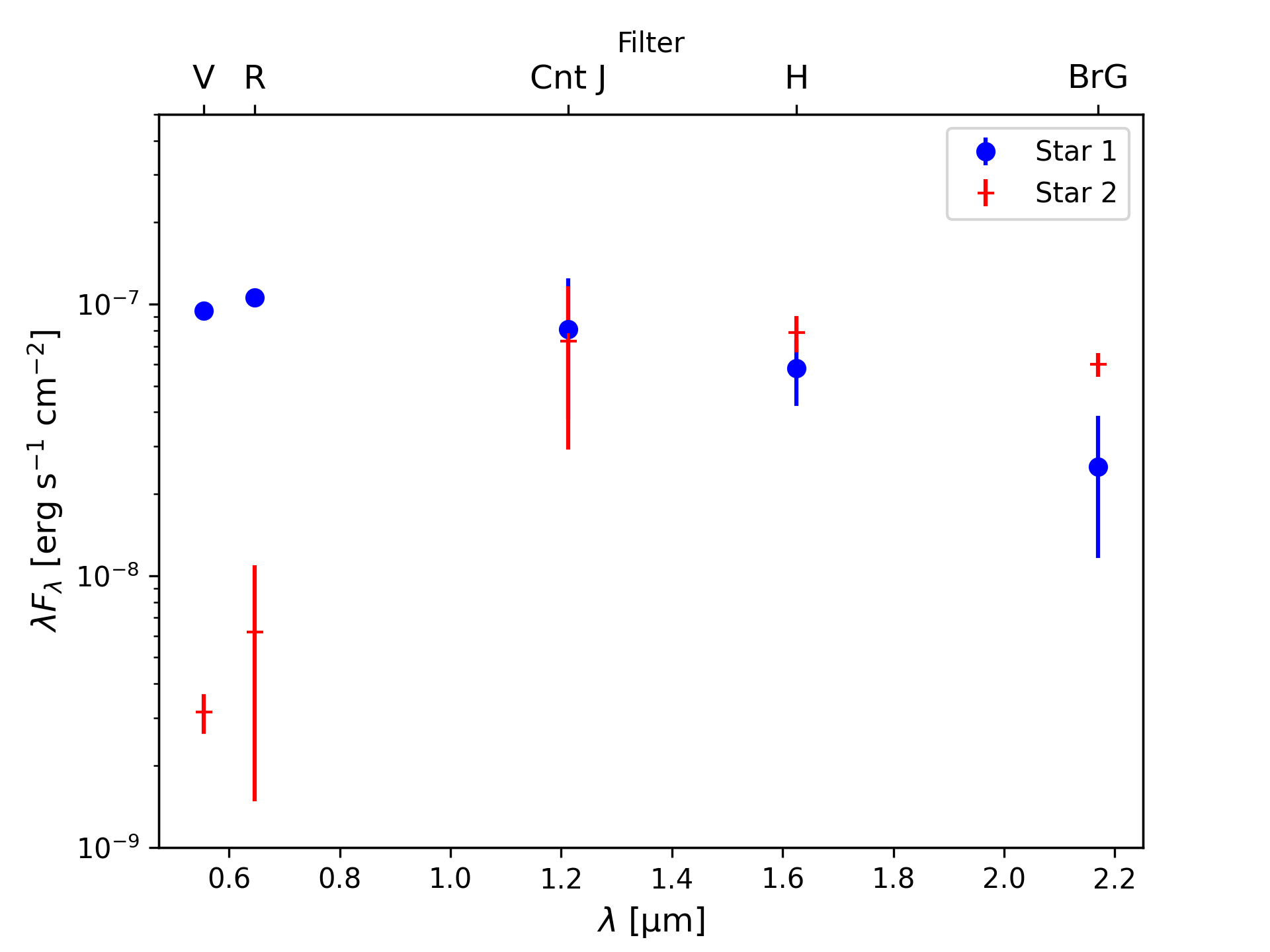}  
    \caption{SED of both components of system, as calculated by extracting photometry from the two stars, resolved in our SPHERE images. Larger uncertainties on the CntJ filter are due to the strong influence of the surrounding nebula. }
    \label{fig:photo}
\end{figure}

Adopting the standard value of $R_V = 3.1$ for the Milky Way interstellar medium, the total extinction reported by Gaia corresponds to $E(\mathrm{B}-\V) = 1.30 \pm 0.01$ mag. This value is consistent with that reported by \bluecite{Molster1999}, considering that they found an extinction due to the circumstellar material of \textcolor{\change}{$E(\mathrm{B}-\V) = 0.22 \pm 0.05$ mag}. This leads to a total extinction in agreement with the one used by \bluecite{Molster1999}. In the following analysis, we adopt the same interstellar reddening: E(B-V) = 1.07 mag.
The measurements of the photometry are presented in Tab.~\ref{tab:photo} and plotted in Fig. \ref{fig:photo}.

\begin{table}[h!]
\centering
\caption{\textcolor{\change}{Photometric measurements for AFGL 4106.}}
\begin{tabular*}{0.482\textwidth}{@{\extracolsep{\fill}}cccc}
\hline
SPHERE Filter & $\lambda_0$ ($\mu$m) & $F_1/F_2$ & $F_1 + F_2$\tablefootmark{1} \\
\hline
V & 0.554 $\pm$ 0.086 & 30.67 $\pm$ 1.51 & 0.51 $\pm$ 0.05\\
NR & 0.646$\pm$ 0.057 & 17 $\pm$ 0.29 & 1.04 $\pm$ 0.50\\
CntJ & 1.213 $\pm$ 0.017 & 1.19 $\pm$ 6.19 & 6.74 $\pm$ 2.82\\
H & 1.625 $\pm$ 0.290 & 0.74 $\pm$ 0.08 & 8.14 $\pm$ 2.69\\
BrG & 2.170 $\pm$ 0.031 & 0.42 $\pm$ 0.05 & 6.05 $\pm$ 1.90\\
\end{tabular*}
\tablefoot{Fluxes are unreddened. The CntJ filter shows particularly high uncertainty due to the strong influence of the surrounding nebula.}
\tablefoottext{1}{$ 10^{-8}$ erg/s/cm$^2$}
\label{tab:photo}
\end{table}

For the first time, we resolved the binary system in the core of AFGL 4106 at multiple wavelengths in the optical and near-infrared, allowing us to build SEDs for both objects. The five photometric data points clearly indicate that the primary star is hotter and more \textcolor{\change}{luminous} than its companion.
To have a better estimation of the stellar properties, we performed radiative-transfer modelling of the system and its surrounding dust shell, as described in the next section.

\subsection{Star properties}
\label{sec:Star_prop}
Figure \ref{fig:photo} give us an overview of the SED of each component of the system in five bands. We used the 1D radiative transfer code DUSTY (Version 4, \bluecite{Ivezic1997}) to model the SED of the system, adopting the dust composition identified by \bluecite{Molster1999}. The model aims to fit the photometric data obtained from VizieR,\footnote[2]{https://vizier.cds.unistra.fr} as well as photometric points constructed by convolving the ISO-SWS spectrum with the ISO filter profiles \tref{table:Data_SED}.  We used the standard MRN dust size distribution \bluecitep{MRN1977} and opted for the solution for radiatively driven winds. The best fit is determined with a minimisation of the $\chi^2$ using a Markov chain Monte Carlo method. We created synthetic fluxes by convolving the model spectrum with the filter profiles of each observational instrument. We then used the $\chi^2$ defined by \bluecite{Yang2023}\textcolor{\change}{:}

\begin{equation}
    \hspace{1cm}\chi^2=\frac{1}{\mathrm{N-p-1}}\sum \frac{[1-f(\lambda,\mathrm{model})/f(\lambda,\mathrm{Obs})]^2}{f(\lambda,\mathrm{model})/f(\lambda,\mathrm{Obs})}
,\end{equation} 

\noindent \textcolor{\change}{where} $f$($\lambda$) = F($\lambda$)/F(K) is the flux at a specific wavelength, $\lambda,$ normalised to the flux in the K band; N is the number of photometric data points \textcolor{\change}{(5 in the case of Tab.~\ref{tab:photo} and 47 in the case of Tab.~\ref{table:Data_SED})}; and p is the number of free parameters (5 to 7). This modified $\chi^2$ balances the contributions from both optical and infrared fluxes by effectively flattening the spectrum. In contrast, a standard $\chi^2$ tends to underweight the infrared data due to the steep decline in flux at longer wavelengths of the Planck function.

\begin{figure}[h]
    \centering
    \includegraphics[width=\linewidth]{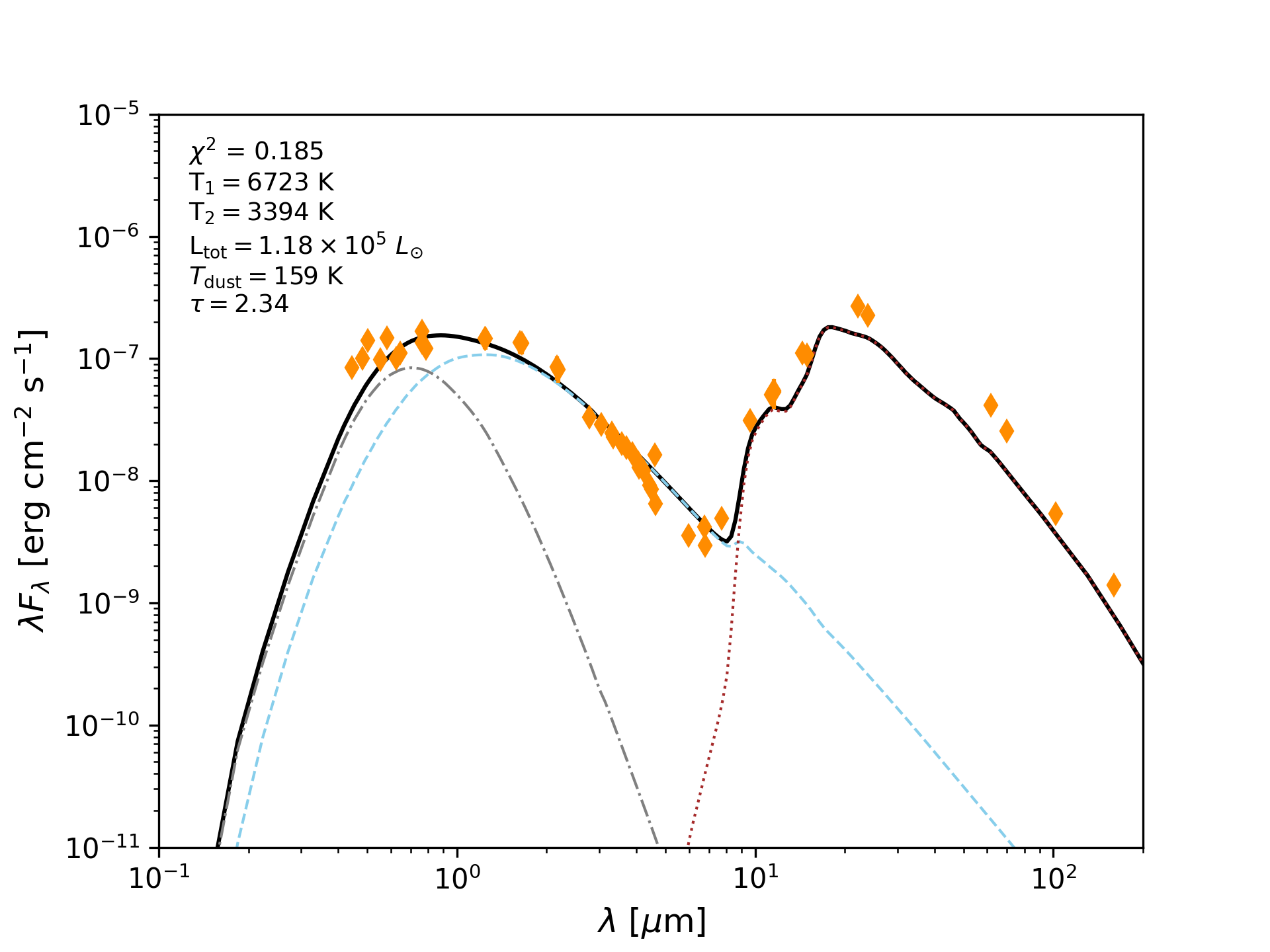}
    \caption{SED fitting using DUSTY version 4.0 at a distance of 3.19 kpc. The orange diamonds show
the photometric measurements, and the black line is the best-fit model consisting of the attenuated flux (dashed light blue), the scattered flux (dot-dashed grey), and dust emission (dotted brown).}
    \label{fig:mod_sed}
\end{figure}

We initially used the SEDs of each component, as shown in Fig.~\ref{fig:photo}, to estimate the stellar temperatures through a first MCMC run. Based on these results, we then constrained the temperatures to narrow intervals centred on the initial estimates. Using the full photometric dataset available from VizieR, we subsequently refined the model to determine the properties of the circumstellar shell and the relative luminosities of the two components in a second MCMC run.\footnote{To facilitate the reusability of our method, we developed DustyPY \url{https://github.com/gtomass/DustyPY.git}, a Python library available on GitHub.} The final results are presented in Fig. \ref{fig:mod_sed} and the model \textcolor{\change}{parameters} in Table. \ref{tab:DUSTY_star}.

\begin{table}[h]
\centering
\caption{\textcolor{\change}{Results of SED fitting.}}
\begin{tabular*}{0.482\textwidth}{@{\extracolsep{\fill}}lcr}
\hline
\multicolumn{3}{c}{Stellar properties}\\ \hline
T$_1$ & &\Tufit  $\pm$ \Tufite K \\
L$_1$ & &(7.9 $\pm$ 0.18) $\times 10^4$  L$_\odot$ \\
R$_1$ & & 209 $\pm$ 12 R$_\odot$ \\
T$_2$ & &\Tdfit $\pm$ \Tdfite K \\
L$_2$ & &(3.8 $\pm$ 0.11) $\times 10^4$ L$_\odot$ \\
R$_2$ & & 567 $\pm$ 99 R$_\odot$ \\ 
\end{tabular*}
\begin{tabular*}{0.482\textwidth}{@{\extracolsep{\fill}}lcr}
 \hline
\multicolumn{3}{c}{Dust Properties}\\ 
    \hline
Grain size ($\mu$m) &  & 0.005-0.25 \\
Optical depth at 0.55 $\mu$m &  & 2.17 $\pm$ 0.14 \\
Dust Temperature & & 158 $\pm$ 8 K\\
Shell inner radius $r_\mathrm{in}$ & & 6.46$\times 10^{14}$ m \\
Shell thickness $r_\mathrm{out}/r_\mathrm{in}$ & & $\sim$40 \\
\end{tabular*}
\begin{tabular*}{0.482\textwidth}{@{\extracolsep{\fill}}lcr}
 \hline
\multicolumn{3}{c}{Dust Composition}\\ 
    \hline
Mass fraction of Mg$_2$SiO$_4$ &  & 4\% \\
Mass fraction of MgSiO$_3$ &  & 10\% \\
Mass fraction of H$_2$O &  & 5\% \\
Mass fraction of $\alpha$-Al$_2$O$_3$ &  & 15\% \\
Mass fraction of MgFeSiO$_4$ &  & 65\% \\
Mass fraction of FeO &  & 1\%
\end{tabular*}
\tablefoot{The SED fitting was done using DUSTY version 4.0 at a distance of 3.19 kpc.}
\label{tab:DUSTY_star}
\end{table}

With such luminosities and temperatures, those stars have radii of R$_1$ = (209 $\pm$ 12) R$_\odot$ and R$_2$ = (567 $\pm$ 99) R$_\odot$ at 3.19 kpc. This indicates that these stars are both supergiants in two different stages of evolution. The primary seems to be the more massive, and the nebula was formed when it expelled its outer layers, leaving the star in the post-RSG phase. The companion is likely in a less evolved stage, with a  radius similar to giant stars, and may be surrounded by a possible expanding shell caused by wind at the surface.

DUSTY computes both the inner radius of the dust shell and an estimate of the mass-loss rate. For our system, the inner radius is calculated as $r_\mathrm{int} = 6.46 \times 10^{14}$ m, corresponding to 1.35 arcsec at a distance of 3.19 kpc. This radius is slightly less than the 1.6-1.9 arcsec derived by \bluecite{vanLoon1999}. To estimate the mass-loss rate, we applied the rescaling relation for dust-driven winds described by \bluecite{vanLoon2000}:

\begin{equation}
    \hspace{2cm}\dot M = \dot M_{\mathrm{DUSTY}} \left(\frac{L}{10^4 L_\odot}\right)^{3/4} \left(\frac{r_{gd}}{200} \frac{\rho_d}{3}\right)^{1/2}
,\end{equation}where $r_{gd} = 100$ is the gas-to-dust ratio assumed by \bluecite{Molster1999} and  $\rho_d = 3.3 \, \mathrm{g \, cm^{-3}} $ represents the bulk density of the dust. Using these parameters, we derive a total mass-loss rate of \textcolor{\change}{$\dot{M} \simeq 4 \times 10^{-3}\, \left(\frac{d}{3.19 \, \mathrm{kpc}}\right)^{1.5} \, M_\odot \, \mathrm{yr^{-1}}.$}

\section{Discussion}
\label{sec:Discussion}
\subsection{Distance to AFGL 4106}
\label{sec:Distance}
The distance to AFGL 4106 is debated. Gaia's parallax gives a distance of 0.94 $\pm$ 0.26 kpc \bluecitep{Gaia2016,Gaia2023}, but when applying the Bailer-Jones method \bluecitep{Bailer-Jones2021}, using a probabilistic approach to estimate stellar distances and a prior constructed from a three-dimensional model of our Galaxy, we find a distance of 1.47$_{-0.4}^{+0.59}$ kpc. Due to its proximity with the Carina nebula \bluecite{Molster1999} assumed that AFGL\,4106 was at a distance of 2.61 kpc 

To improve the distance estimation, we combined the interstellar extinction measurement from \bluecite{Molster1999} with the value derived from GSP-Spec data (de Laverny, in preparation, private communication), and compared them with the extinction maps from G-TOMO \bluecitep{Lallement2022, Vergely2022}. G-TOMO delivers three-dimensional extinction maps within a 5 $\times$ 5 $\times$ 0.8 pc region centred on the Sun, using Gaia EDR3 and 2MASS photometry and parallaxes. This provides access to distance-dependent extinction curves in the solar neighbourhood, which can then be inverted: given the extinction, we can estimate a star's distance. Taking an extinction value of \textcolor{\change}{$E(\mathrm{B}-\V) = 1.07 \pm 0.2$ mag}, we derive a distance of 3.19$_{-0.19}^{+0.45}$ kpc.

\begin{figure}[h]
    \centering
    \includegraphics[width=\linewidth]{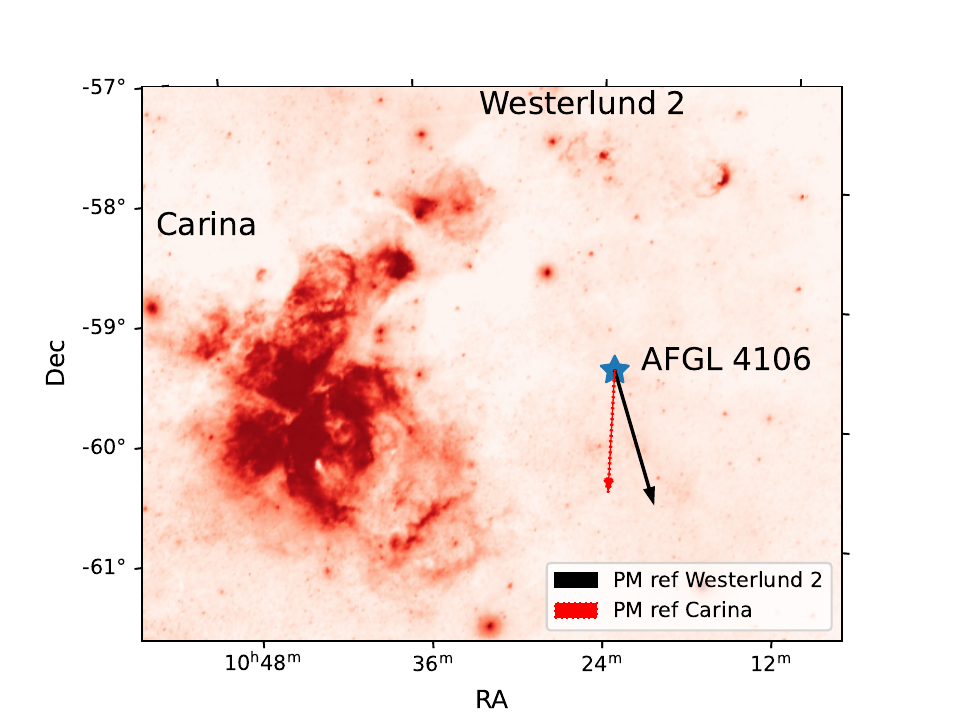} 
    \caption{Relative proper motion of AFGL 4106 with the Carina (red arrow) and the open cluster Westerlund 2 (black arrow).}
    \label{fig:Rel-PM}
\end{figure}

As these are massive stars, their lifetime is particularly short - around 11 Myr for a 15 M$_\odot$ star - so they are necessarily located close to \textcolor{\change}{the} cluster where they were formed. The two closest star forming regions are the Carina nebula and the open cluster Westerlund 2, which are, respectively, at a projected distance of $\sim$2.1° and $\sim$1.8° of AFGL 4106 and at a distance of 2.61 kpc and 4.13 kpc. In the Carina reference frame, the proper motion is oriented in the direction of the south, with an absolute value of 4 \textcolor{\change}{mas yr$^{-1}$} \bluecitep{Gaia2023}. However, the proper motion of AFGL 4106 is not consistent with an origin in the Carina Nebula, as shown in \fref{fig:Rel-PM}. When re-referenced to the Carina reference frame, the proper motion vector does not point towards the nebula, indicating a different origin or kinematic history. In the Westerlund referential frame, the proper motion is also directed towards the south, with an absolute value of 4.5 mas yr$^{-1}$. However, in this frame, the motion appears to be moving directly away from the cluster. 

\subsection{Radial profiles}
\label{sec:Radial}

\begin{figure}[h!]
    \centering
    \includegraphics[width=\linewidth]{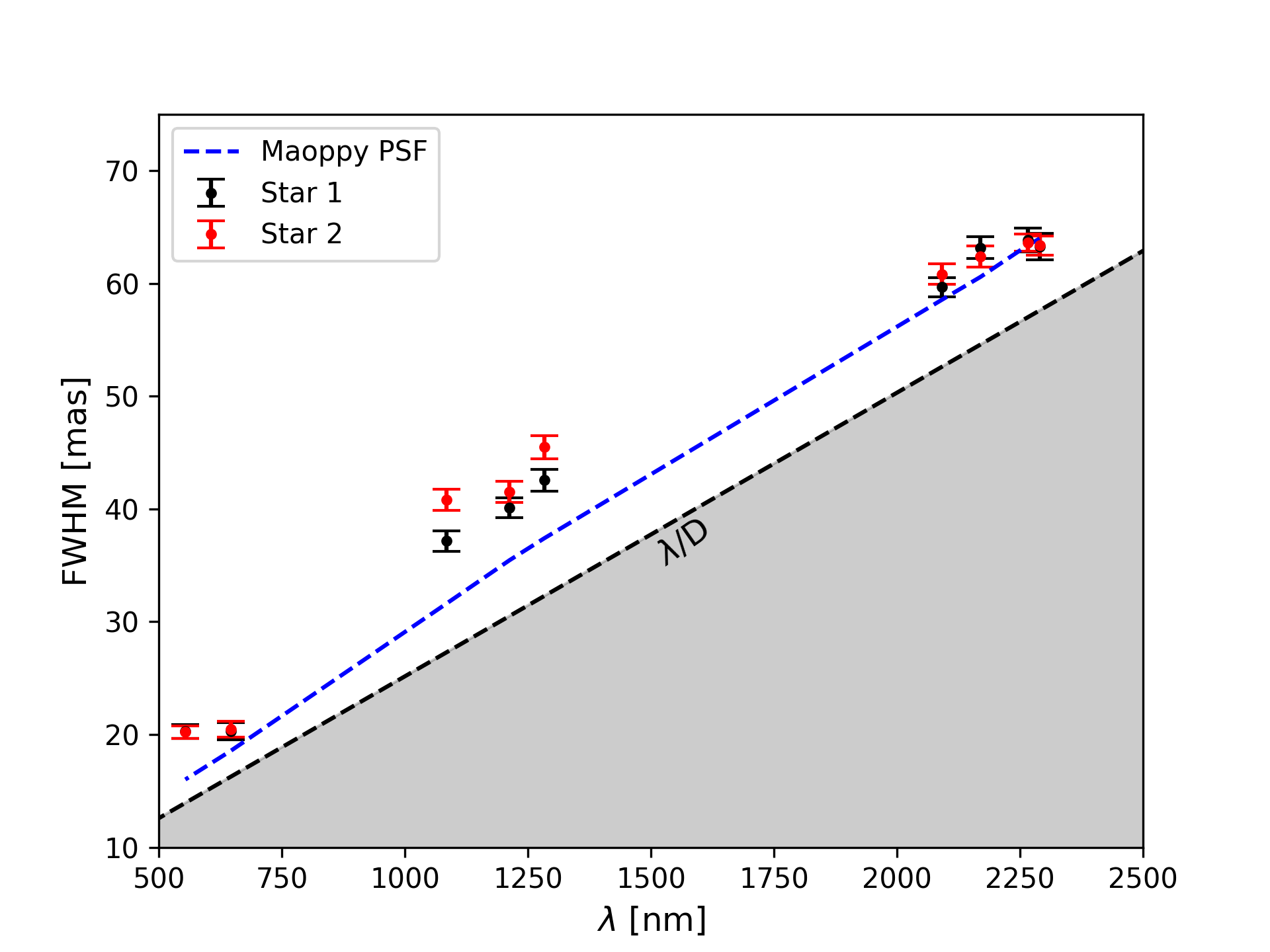}
    \caption{Radial \textcolor{\change}{profile} of both stars of AFGL 4106 at different wavelengths.}
    \label{fig:rad}
\end{figure}

The radial profiles of both stars were computed from ZIMPOL and IRDIS images, and the full width at half maximum (FWHM) was derived for each filter \fref{fig:rad}. The measured FWHM values were then compared to the diffraction limit, $\lambda/D$ (black line), and to the IRDIS and ZIMPOL-modelled point spread function (PSF) generated using the Maoppy code \bluecitep{Fetick2019} (blue line).

In each filter, both stars appear resolved. However, despite the distance to AFGL 4106, the photosphere of the companion (R$_2$ = 567 $\times$ $\frac{\mathrm{d}}{3.19 \text{ kpc}}$ R$_\odot$ $\sim$0.82 mas) cannot be resolved with our observations. This indicates that the resolved structure is likely a circumstellar shell. While the exact size of this shell is challenging to determine, the FWHM at longer wavelengths tends to align with the modelled PSF. This suggests a shell size of approximately 65 mas. At a distance of 3.19kpc, this corresponds to an envelope size of roughly 208 au ($\sim$80 R$_2$) .

\subsection{Shaping and mass}
\subsubsection{HR diagram}
\label{Sec:HR}
\begin{figure}[h!]
    \centering
    \includegraphics[width=\linewidth]{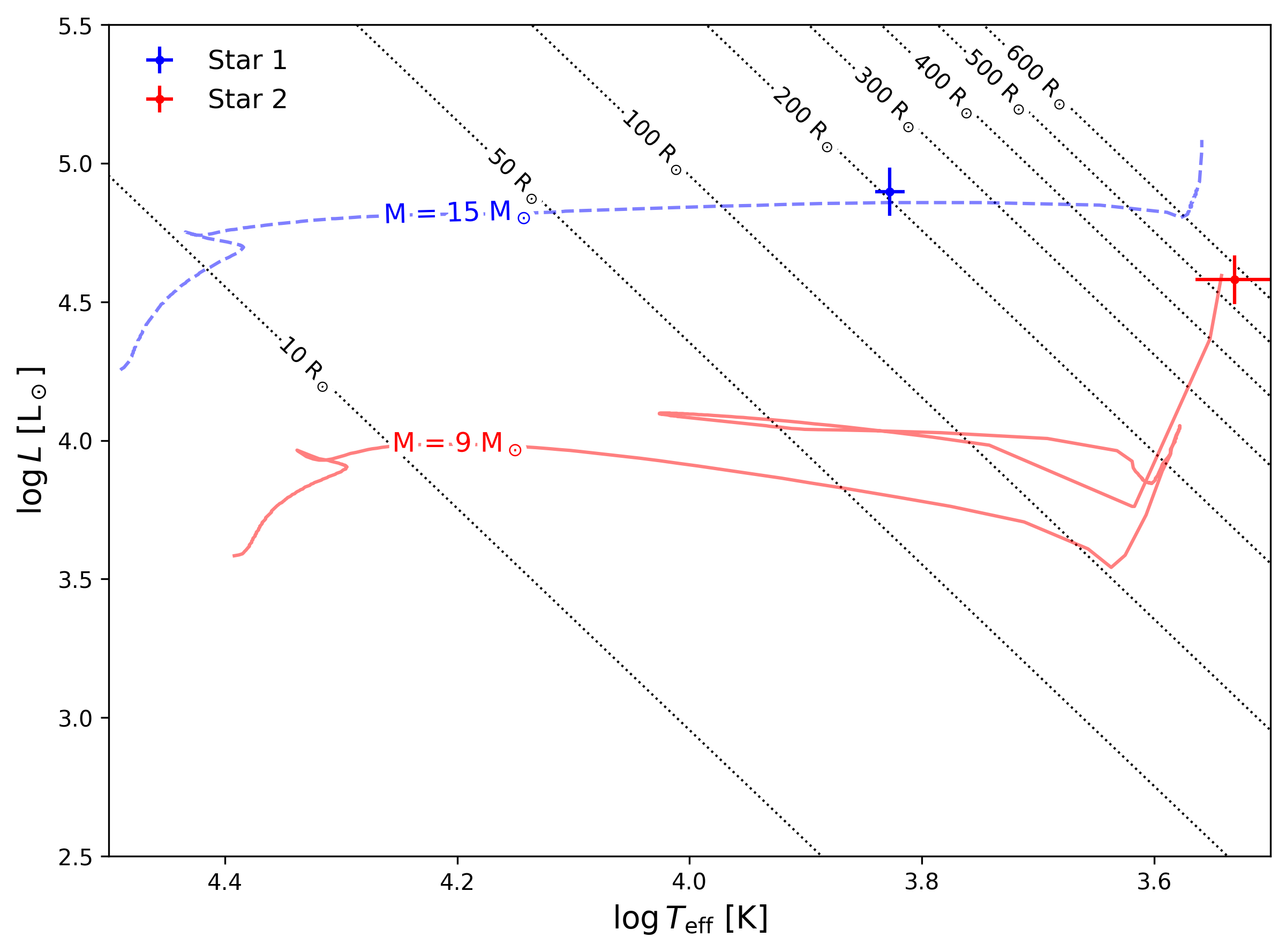}
    \caption{HR diagram of two components of AFGL 4106 at 3.19 kpc. Evolutionary tracks are from \bluecite{Ekstrom2012}. Isoradius \textcolor{\change}{lines are} overplotted with dotted lines.}
    \label{fig:ET}
\end{figure}

Based on the SED fitting results, we placed both components of the system on a Hertzsprung-Russell diagram \fref{fig:ET}. Their positions were then compared with the evolutionary tracks from \bluecite{Ekstrom2012}, considering both rotating and non-rotating stellar models. At a distance of 3.19 kpc, both stars align well with the theoretical tracks.

The temperature and luminosity of the primary are consistent with those of a rotating star with an initial mass of approximately 15 M$_\odot$. According to the evolutionary tracks, such a star would be around 13.8 Myr old, with a current mass of 14.5 M$_\odot$. However, this inferred age is significantly older than the estimated age of the candidate star-forming region, which is only 2-3 Myr \bluecitep{Zeidler2015}. This discrepancy may indicate either an incorrect assumption about the system's origin or a deviation from standard evolutionary pathways due to binary interaction.

Similarly, the secondary's temperature and luminosity are consistent with those of a non-rotating star with an initial mass of about 9 M$_\odot$, corresponding to an age of roughly 30~Myr and a current mass of 8.7 M$_\odot$. \textcolor{\change}{Once again, this age is incompatible with that of Westerlund~2 and, more importantly, it does not agree with the age inferred for the primary in the previous paragraph. This discrepancy further suggests that standard single-star evolutionary tracks may not adequately describe the system's past.}

\begin{figure}[h!]
    \centering
    \includegraphics[width=\linewidth]{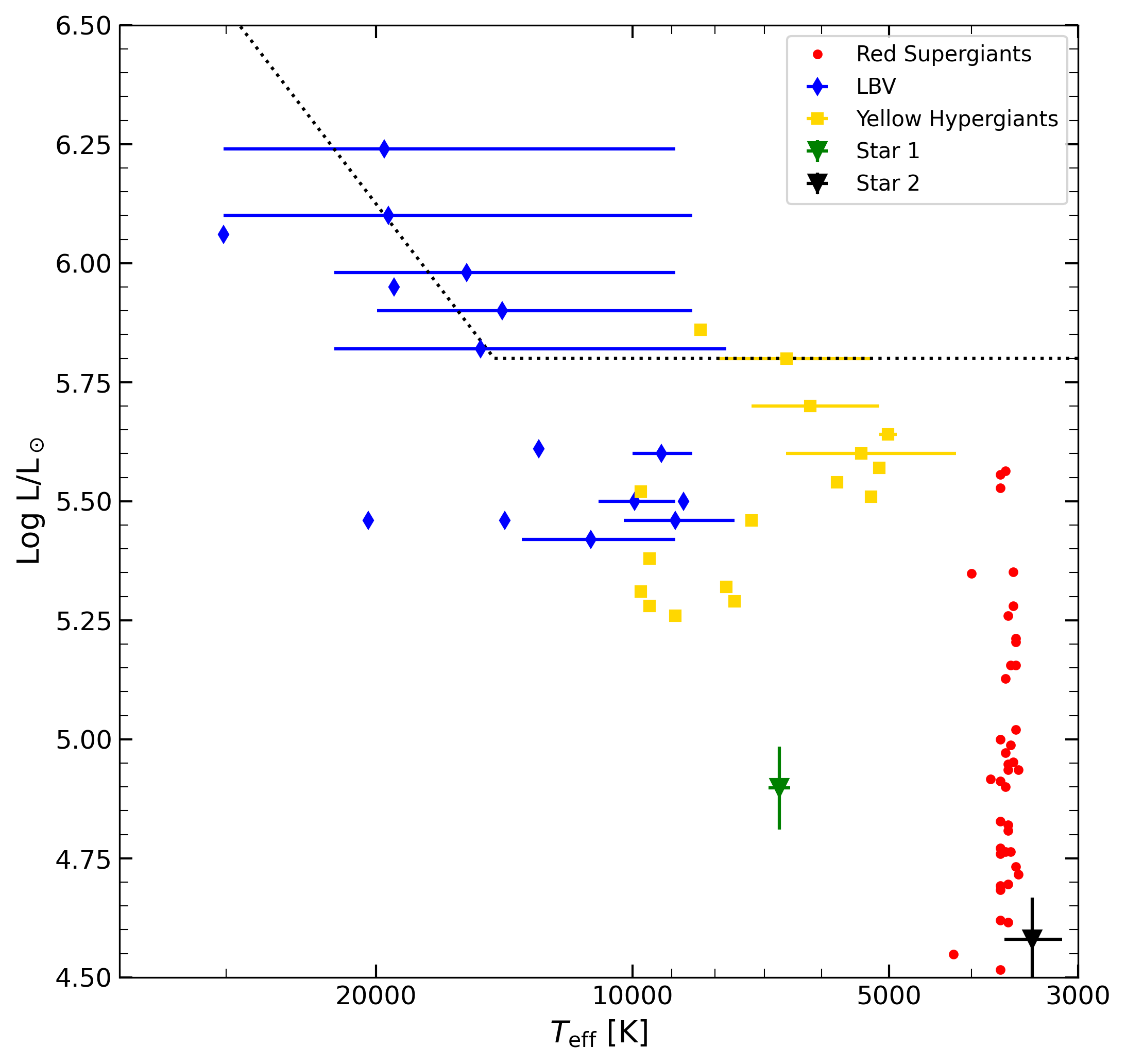}
    \caption{HR diagram of two components of AFGL 4106 at 3.19 kpc. Stars from \bluecite{Oudmaijer2009} are overplotted. The dotted line corresponds to the Humphreys-Davidson limit.}
    \label{fig:HR-comp}
\end{figure}

We also placed both components on a Hertzsprung-Russell diagram \fref{fig:HR-comp} alongside luminous blue variables, yellow hypergiants, and RSGs from \bluecite{Oudmaijer2009}.

The secondary falls clearly within the RSG region, consistent with its classification. In contrast, the primary is located just below the so-called yellow void. It is too hot to be classified as an RSG and too faint to be considered a yellow hypergiant, placing it instead among the yellow supergiants (YSGs). Intermediate-mass stars can experience a 'blue loop' during their evolution, temporarily evolving from the RSG phase to warmer YSG or even blue-supergiant (BSG) phases. However, the primary's mass (M > 12 M$_\odot$) is likely too high for this mechanism to occur under standard evolutionary models.
The system's binarity is a factor that may alter its evolutionary pathway. However, the binary is too wide to have  \textcolor{\change}{affected} the evolution of the primary. For stars to truly affect each other they should be close, at most within a \textcolor{\change}{few astronomical units \bluecitep{Sana2012}.} \textcolor{\change}{This leaves the possibility that AFGL 4106 experienced interaction with a third, close companion, or even a past merger, open, which might help explain a blue-loop-like excursion or its unusually low post-RSG luminosity.}

\subsubsection{Nebular shaping: Influence of the companion}
\label{sec:shaping}
We see in Sect. \ref{sec:Distance} \textcolor{\change}{that} the nebula shows signs of oscillation (Fig. \ref{fig:el}).  
These oscillations might come from the variation of the orbital speed of a companion, alongside the orbit curve \bluecitep{ElMellah2020}. When the ratio $\eta = v_{wind}/v_{orb}$ between the wind speed and orbital velocity is low, the ejecta is constrained in the orbital plane. Thus, depending on the angle of observation, the shape of this ejecta might be elliptical, whereas when $\eta$ is larger, the ejecta is spherical, so the nebula will appear spherical for all lines of sight.

If the observed variation in ellipticity is attributed to the orbital motion of the binary, by applying Kepler's third law with an oscillation period of approximately 1300-1950 years (depending on the assumed wind velocity of 20-30 km s$^{-1}$, Sect.~\ref{sec:IRDIS}), the resulting total system mass is estimated to be in the range of 175-400 M$_\odot$. However, this estimate is inconsistent with the stellar parameters derived in the previous section. Moreover, such high total mass is improbable, as the formation of a binary system comprising two stars each exceeding 80 M$_\odot$ is considered unlikely based on current star formation models.

\textcolor{\change}{The presence of a third object to explain the ellipticity oscillations is plausible but remains unsupported observationally: no such companion is detected in either the imaging or the spectroscopic data. Assuming an unresolved companion at the resolution limit of our observations (20 mas), Kepler's law yields a total mass of only 0.06-0.15\,M$_\odot$, far from the $\sim 15$\,M$_\odot$ estimated in Sect.~\ref{Sec:HR}.}

\textcolor{\change}{\bluecite{vanLoon1999} reported radial-velocity variations with a period of 4500 days, interpreted as the orbital motion of the known binary and implying stellar masses $\leq 15\,\sin^{-3} i$\,M$_\odot$, which is consistent with our results. If, instead, the radial velocity oscillation were attributed to an orbit involving a third companion located at 20 mas, the resulting total mass would be $\sim 1700$\,M$_\odot$, which is unphysical. Matching the mass estimated in Sect.~\ref{Sec:HR} would require a separation smaller than $\sim 4.1$\,mas (i.e. $\sim 13$\,au at 3.19 kpc).}

\textcolor{\change}{Such a close companion could, in principle, exist and might account for the radial-velocity variations reported by \bluecite{vanLoon1999}. However, it cannot explain the large-scale oscillations observed in the nebula, which would require a much longer orbital period. Therefore, while the presence of a close companion cannot be ruled out, it is unlikely to be responsible for the observed nebular morphology.}

The shaping of the nebula is likely the result of a complex dynamic driven by, first, the influence of one or many companions and, second, by a complex mechanism of ejection. It is also important to note that, the semi minor axis is almost aligned with the proper motion of the system. The interaction with the interstellar medium (ISM) may be responsible for the peculiar morphology of the nebula, particularly its large-scale elliptical shape with an eccentricity of approximately 0.5, as the stellar wind is decelerated by the ISM through the formation of a 'shock front'.

\section{Conclusions}
\label{sec:Conclusion}
The main results of this study can be summarised as follows:
\begin{itemize}
    \item AFGL 4106 is a binary system composed of a post-RSG and an RSG with a separation of 0.272 $\pm$ 0.017 arcsec and temperatures T$_1$ = \Tufit $\pm$ \Tufite K and T$_2$ = \Tdfit $\pm$ \Tdfite K.
    \item The primary star appears slightly brighter than its companion, with a bolometric-luminosity ratio of L$_1$/L$_2$ = 2.09, primarily attributed to its temperature being approximately two times higher.
    \item Both stars have high bolometric luminosities, with L$_1$ = (7.9 $\pm$ 0.18) $(\frac{\mathrm{d}}{3.19 \text{ kpc}})^2$ $\times 10^4$ L$_\odot$ and L$_2$ = (3.8 $\pm$ 0.11) $(\frac{\mathrm{d}}{3.19 \text{ kpc}})^2$ $\times 10^4$ L$_\odot$ with d in kpc.
    \item The radii  R$_1$ = (209 $\pm$ 12)$(\frac{\mathrm{d}}{3.19 \text{ kpc}})$ R$_\odot$ and R$_2$ = (567 $\pm$ 99)$(\frac{\mathrm{d}}{3.19 \text{ kpc}})$ R$_\odot$ reveal that the stars are in two close, but different, stages of evolution: the primary in a post-RSG phase and the companion in a RSG phase.
    \item The stellar evolution model gives an estimation of 15 \Msol for the primary and 9 \Msol for the secondary. 
    \item The proximity to the open cluster Westerlund 2 and the relative movement of both systems suggest that the star could have originated from the cluster. This gives an upper bound on the age of the stars $\lesssim$ 2-3 Myr. 
    \item The binary is surrounded by a shell of $\sim$80 R$_2$. 
    \item The complex morphology of the nebula reveals a non-isotropic and non-continuous matter ejection from the system, as well as a strong interaction with the ISM. 
\end{itemize}

This study of AFGL 4106 will enable us to constrain the dust and evolutionary models in this type of system, to better understand the formation and evolution of such nebulae. Further calculation and hydrodynamic models, combined with observations of the dynamics of the nebula, will help us to study the physics of such systems and the final step of star evolution.

\begin{acknowledgements}
This work was supported by the French National Research Agency (ANR) through the PEPPER project (ANR-20-CE31-0002). I would like to thank the referee for their constructive comments, which helped improve the quality of this paper.
\end{acknowledgements}

\appendix
\onecolumn
\section{Observation and Data}
\begin{table*}[h!]
    \centering
    \caption{\textcolor{\change}{SPHERE/VLT Observation log of AFGL 4106.}}
    \begin{tabular*}{\textwidth}{@{\extracolsep{\fill}}cccccccc}
    \hline
        Date & Instrument Mode & Filter & $\lambda$ (nm) & $\Delta\lambda$ (nm) & Coronagraph & DIT (s) & seeing (arcsec) \\ \hline
        2021-04-03 & IRDIS-DBI\tablefootmark{a} & NB PaB & 1283 & 18 & NO & 2.0 & 0.63 \\
        2021-04-03 & IRDIS-DBI & NB HeI & 1085 & 14 & NO & 4.0 & 0.74 \\ 
        2021-04-03 & IRDIS-DBI & NB H2 & 2124 & 31 & NO & 0.84 & 0.69 \\ 
        2021-04-03 & IRDIS-DBI & NB CO & 2290 & 33 & NO & 2.0 & 0.89 \\ 
        2021-04-03 & IRDIS-DBI & NB CntK1 & 2091 & 34 & NO & 4.0 & 0.71 \\ 
        2021-04-03 & IRDIS-DBI & NB CntK2 & 2266 & 32 & NO & 2.0 & 0.74 \\
        2021-04-03 & IRDIS-DBI & NB CntJ & 1213 & 17 & NO & 1.0 & 0.72 \\
        2021-04-03 & IRDIS-DBI & NB BrG & 2170 & 31 & NO & 0.84 & 0.69 \\ 
        2021-04-03 & IRDIS-DBI & BB J & 1245 & 240 & NO & 2.0 & 1.41 \\ 
        2021-04-03 & IRDIS-DBI & BB H & 1625 & 290 & NO & 2.0 & 0.74 \\ 
        2021-04-03 & IRDIS-DBI & BB Ks & 2182 & 300 & NO & 0.84 & 1.21 \\
        2019-03-07 & ZIMPOL P3\tablefootmark{b} & B Ha & 655.6 & 5.5 & YES & 10 & 0.67 \\
        2019-03-07 & ZIMPOL P3 & CntHa & 644.9 & 4.1 & YES & 10 & 0.67 \\
        2019-03-07 & ZIMPOL P3 & V & 554 & 80.6 & NO & 1.2 & 0.35 \\
        2019-03-07 & ZIMPOL P3 & NR & 645.9 & 56.7 & NO & 1.2 & 0.35 \\ 
        2019-03-07 & ZIMPOL P3 & I PRIM & 789.7 & 152.7 & YES & 10 & 1.2 \\ \hline
    \end{tabular*}
    \label{tab:log}
    \tablefoottext{a}{DBI (Dual Band Imaging) mode use two narrow band filter simultaneously.}
    \tablefoottext{b}{P3 (Polarimetric mode) use a polarizer and a coronagraph.}
\end{table*}

\begin{table*}[h!]
    \centering
    \caption{\textcolor{\change}{VizieR photometric data points used in the SED fitting.}}
    \begin{tabular*}{\textwidth}{@{\extracolsep{\fill}}ccccp{5.5cm}}
        \hline
        Wavelength ($\mu$m) & Flux (10$^{-8}$ erg/s/cm$^{2}$) & Flux error (10$^{-8}$ erg/s/cm$^{2}$) & Filter & Catalog \\
        \hline
       0.444 &  0.1600   & 0.0027& Johnson:B & GSC2.3, SPM4, UCAC4-RPM, AAVSO \\
        0.482 &  0.2737   & -     & SDSS:g &  Gaia DR3 synphot \\
        0.504 &  0.4733   & 0.0030& GAIA/GAIA3:Gbp &  Gaia EDR3/DR3 \\
        0.554 &  0.5070   & -     & Johnson:V & SPM4, HYP/TYC,  UCAC4-RPM \\
        0.582 &  0.9371   & 0.0051& GAIA/GAIA3:G & Gaia DR3/EDR3 \\
        0.625 &  0.8206   & -     & SDSS:r & Gaia DR3 synphot \\
        0.647 &  1.0334   & -     & Cousins:R & Gaia DR3 synphot \\
        0.762 &  2.1676   & 0.0197& GAIA/GAIA3:Grp & Gaia DR3/EDR3 \\
        0.763 &  2.6858   & -     & SDSS:i & Gaia DR3 synphot \\
        0.789 &  2.1517   & -     & Cousins:I & Gaia DR3 synphot \\
        1.239 &  6.5954   & 1.2098& 2MASS:J & 2MASS, IRAS, UCAC4, HSOY, MDFC, PPMXL, WISE, NOMAD, ASCC, UCAC5, ATLAS-REFCAT2, SPM4 \\
        1.250 &  6.6593   & 1.2391& Johnson:J & Tycho-2, JSDC, 2MASS \\
        1.630 &  8.1170   & 1.5695& Johnson:H & JSDC, 2MASS, Tycho-2 \\
        1.649 &  8.0712   & 1.5656& 2MASS:H & 2MASS, WISE, ATLAS-REFACT2, NOMAD, HSOY, UCAC4, IRAS, ASCC, SPM4, UCAC5, MDFC, PPMXL \\
        2.164 &  6.0581   & 1.3356& 2MASS:Ks & 2MASS, ASCC, MDFC, PPMXL, SPM4, WISE, UCAC4, IRAS \\
        2.190 &  5.8041   & 1.2776& Johnson:K & 2MASS, Tycho-2, JSDC \\
        2.776 &  2.6022   & 0.0002& IS0\_SW4 & ISO SWS \\
        3.042 &  2.3349   & 0.0002& IS0\_SW7 & ISO SWS \\
        3.302 &  2.0181   & 0.0002& IS0\_SW2 & ISO SWS \\
        3.350 &  1.8860   & 0.0060& WISE:W1 & unWISE, IRAS, WISE \\
        3.574 &  1.6879   & 0.0005& IS0\_SW1 & ISO SWS \\
        3.697 &  1.5728   & 0.0002& IS0\_SW6 & ISO SWS \\
        3.872 &  1.4151   & 0.0003& IS0\_SW9 & ISO SWS \\
        4.052 &  1.2077   & 0.0002& IS0\_SW8 & ISO SWS \\
        4.074 &  1.1064   & 0.0002& IS0\_SW5 & ISO SWS \\
        4.219 &  1.0399   & 0.0002& IS0\_SW11 & ISO SWS \\
        4.423 &  0.7979   & 0.0001& IS0\_SW3 & ISO SWS \\
        4.485 &  0.7442   & 0.0001& IS0\_LW1 & ISO SWS \\
        4.600 &  1.4256   & 0.0608& WISE:W2 &  WISE, unWISE, IRAS \\
        4.628 &  0.5648   & 0.0001& IS0\_SW10 & ISO SWS \\
        5.993 &  0.3230   & 0.0001& IS0\_LW4 & ISO SWS \\
        6.749 &  0.3857   & 0.0001& IS0\_LW2 & ISO SWS \\
        6.784 &  0.2712   & 0.0000& IS0\_LW5 & ISO SWS \\
        7.734 &  0.4565   & 0.0001& IS0\_LW6 & ISO SWS \\
        9.641 &  2.9286   & 0.0008& IS0\_LW7 & ISO SWS \\
        11.316 & 4.8278  & 0.0004  & IS0\_LW8 & ISO SWS \\
        11.501 & 5.1334  & 0.0019  & IS0\_LW10 & ISO SWS \\
        11.560 & 5.0831  & 1.4264  & WISE:W3 & WISE \\
        11.590 & 5.1991  & 0.2587  & IRAS:12 & IRAS \\
        14.394 & 10.7475 & 0.0045  & IS0\_LW3 & ISO SWS \\
        14.897 & 10.3227 & 0.0014  & IS0\_LW9 & ISO SWS \\
        22.091 & 26.1920 & -       & WISE:W4 & WISE \\
        23.880 & 22.0113 & 0.8788  & IRAS:25 & IRAS \\
        61.850 & 4.1297  & 0.4944  & IRAS:60 & IRAS \\
        69.999 & 2.5183  & 0.0043  & Herschel/PACS:70 & Herschel/PACS \\
        101.949 & 0.5322 & 0.0588   & IRAS:100 & IRAS \\
        160.000 & 0.1396 & 0.0015   & Herschel/PACS:160 & Herschel/PACS \\
        \hline
        \end{tabular*}
    \label{table:Data_SED}
    \tablefoot{In cases where multiple measurements were available at the same wavelength, the mean flux value was used.}
\end{table*}
\twocolumn
\section{SPHERE Images}

\begin{figure}[h!]
    \centering
    \includegraphics[width=\linewidth]{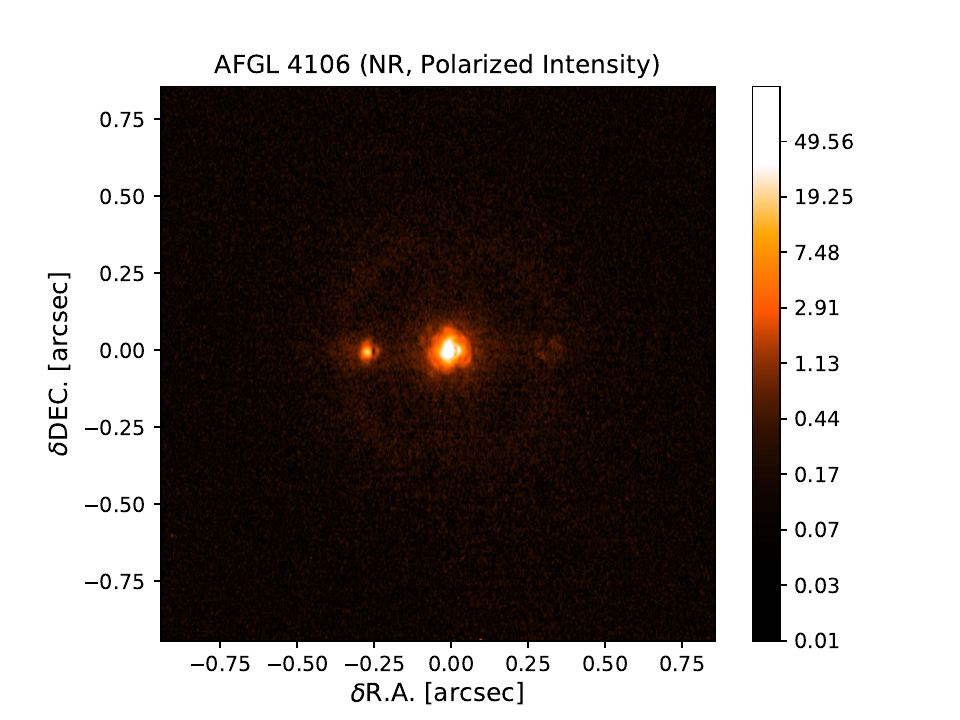}  
    \caption{ZIMPOL Polarized intensity map of AFGL 4106 in filter N\_R.}
    \label{fig:NR}
\end{figure}

\begin{figure}[h!]
    \centering
    \includegraphics[width=\linewidth]{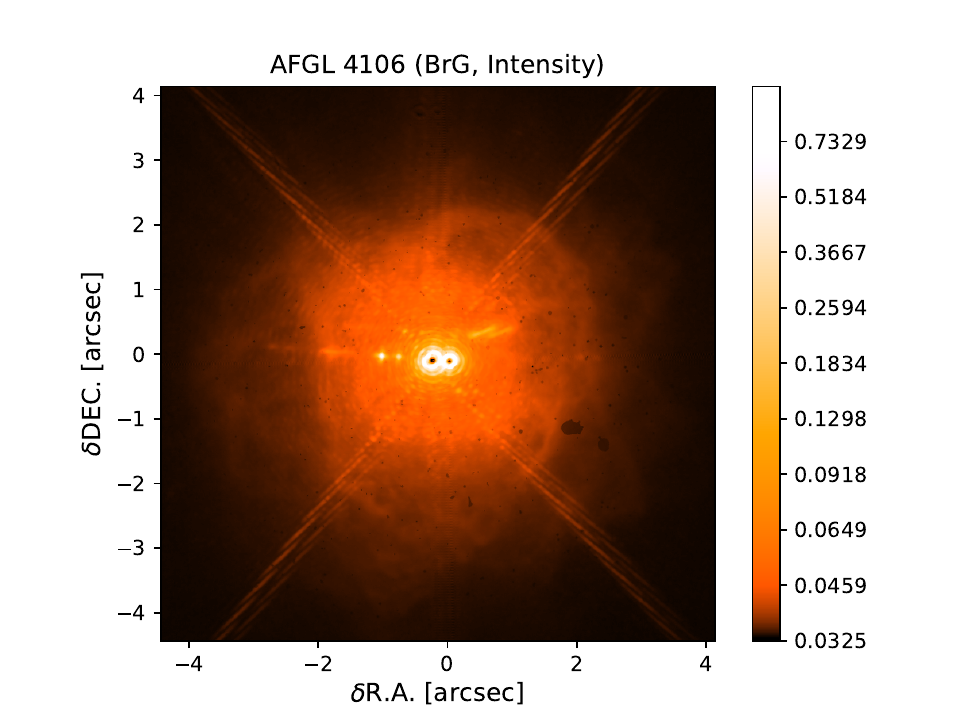}  
    \caption{IRDIS intensity map of AFGL 4106 in filter BrG.}
    \label{fig:BrG}
\end{figure}

\begin{figure}[h!]
    \centering
    \includegraphics[width=\linewidth]{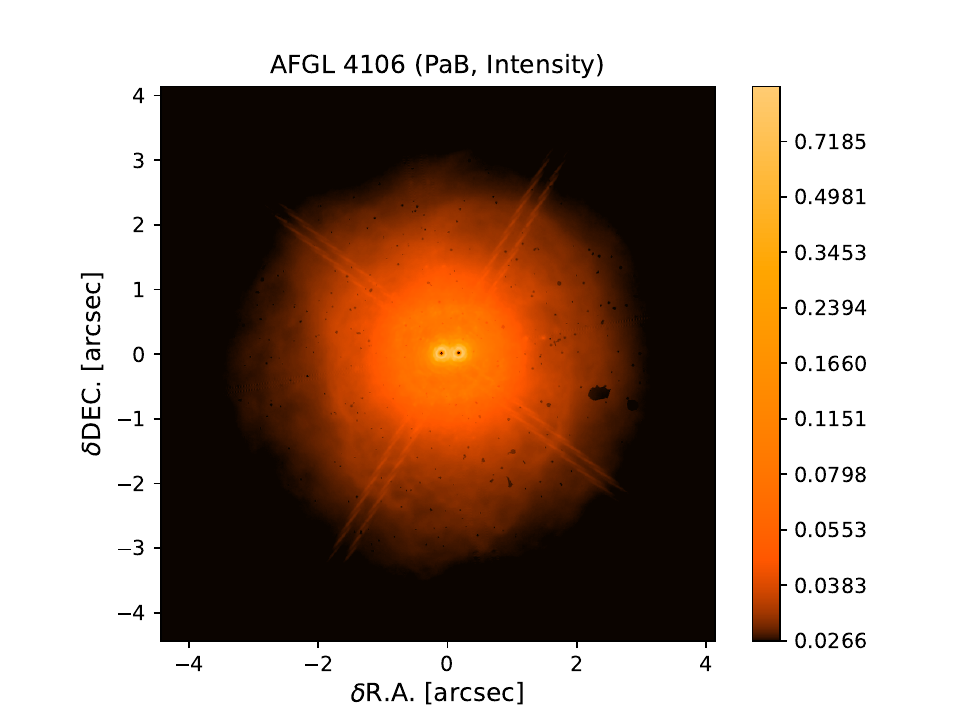}  
    \caption{IRDIS intensity map of AFGL 4106 in filter PaB.}
    \label{fig:PaB}
\end{figure}

\begin{figure}[h!]
    \centering
    \includegraphics[width=\linewidth]{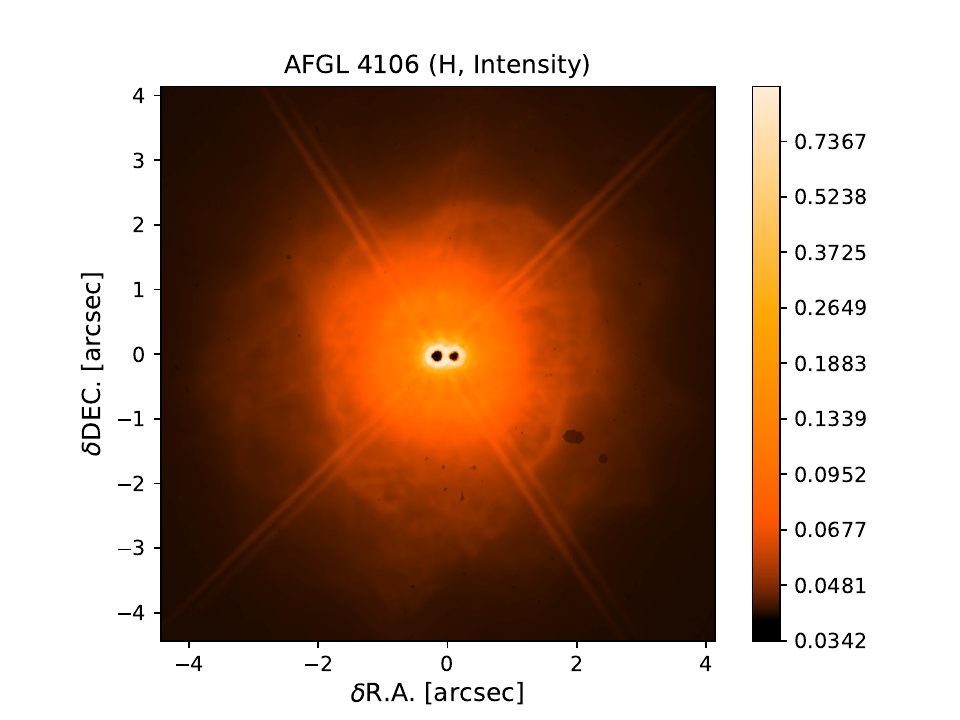}  
    \caption{IRDIS intensity map of AFGL 4106 in filter H.}
    \label{fig:H}
\end{figure}

\begin{figure}[h!]
    \centering
    \includegraphics[width=\linewidth]{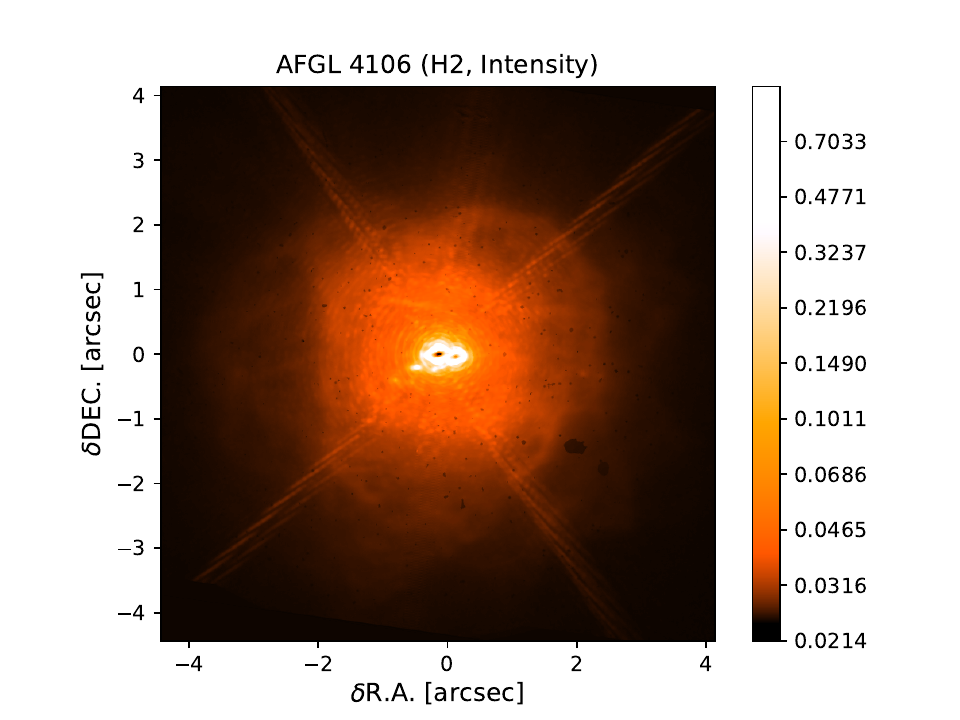}  
    \caption{IRDIS intensity map of AFGL 4106 in filter H2.}
    \label{fig:H2}
\end{figure}

\begin{figure}[h!]
    \centering
    \includegraphics[width=\linewidth]{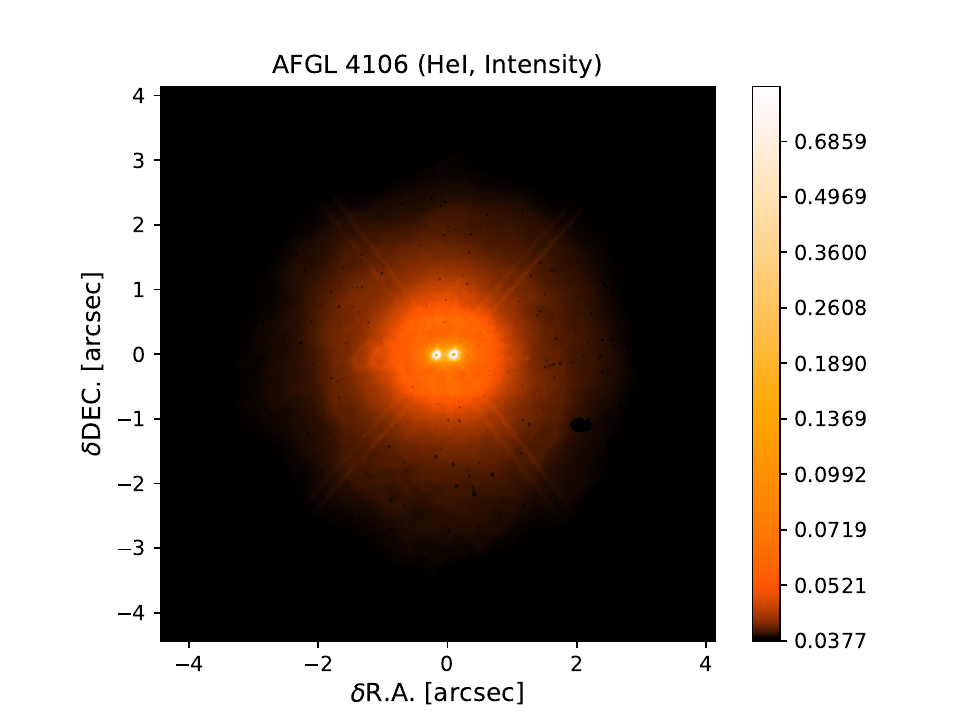}  
    \caption{IRDIS intensity map of AFGL 4106 in filter HeI.}
    \label{fig:HeI}
\end{figure}

\begin{figure}[h!]
    \centering
    \includegraphics[width=\linewidth]{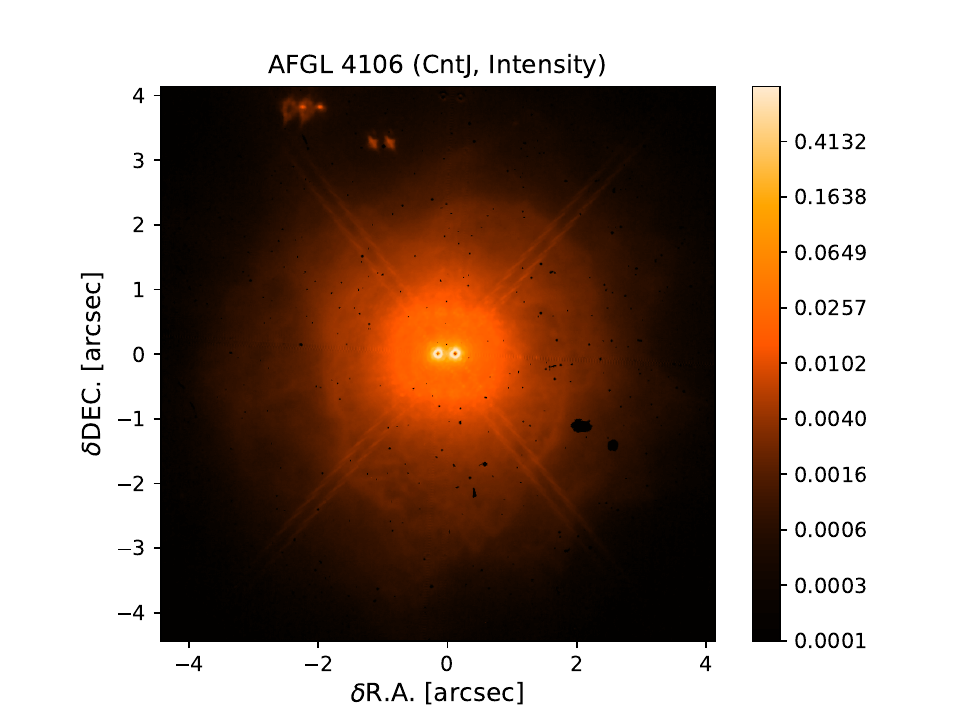}  
    \caption{IRDIS intensity map of AFGL 4106 in filter CntJ.}
    \label{fig:CntJ}
\end{figure}

\begin{figure}[h!]
    \centering
    \includegraphics[width=\linewidth]{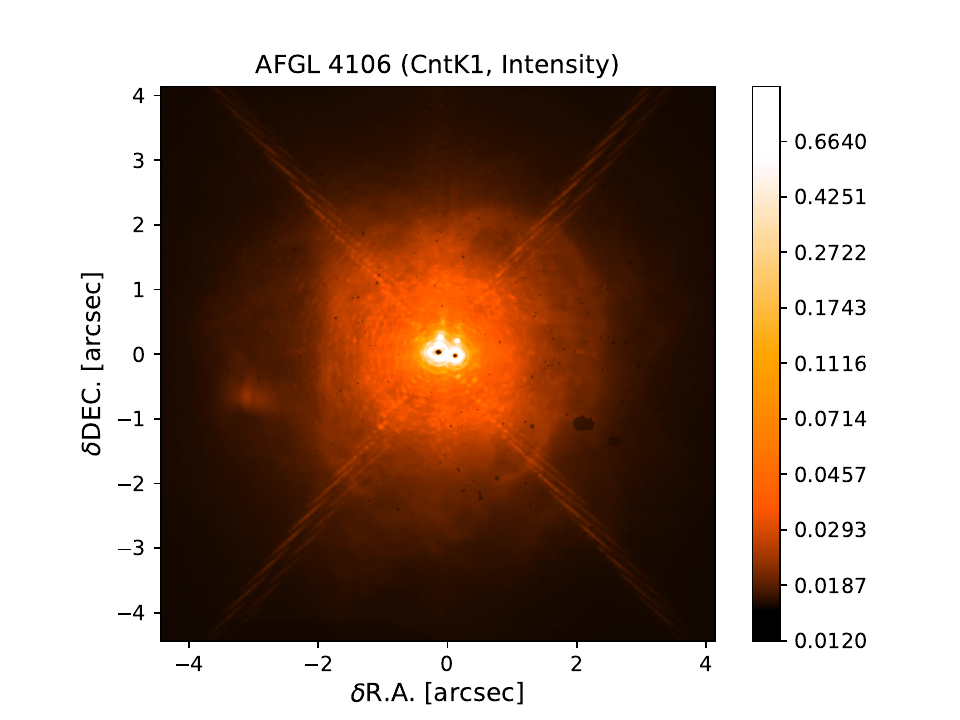}  
    \caption{IRDIS intensity map of AFGL 4106 in filter CntK1.}
    \label{fig:CntK1}
\end{figure}

\begin{figure}[h!]
    \centering
    \includegraphics[width=\linewidth]{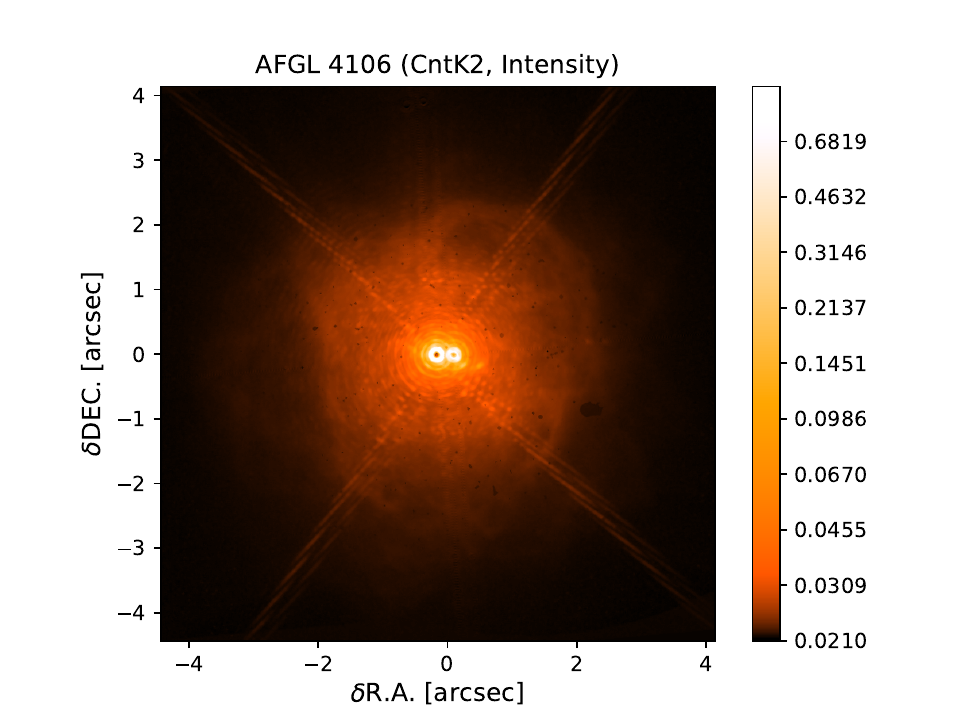}  
    \caption{IRDIS intensity map of AFGL 4106 in filter CntK2.}
    \label{fig:CntK2}
\end{figure}

\begin{figure}[h!]
    \centering
    \includegraphics[width=\linewidth]{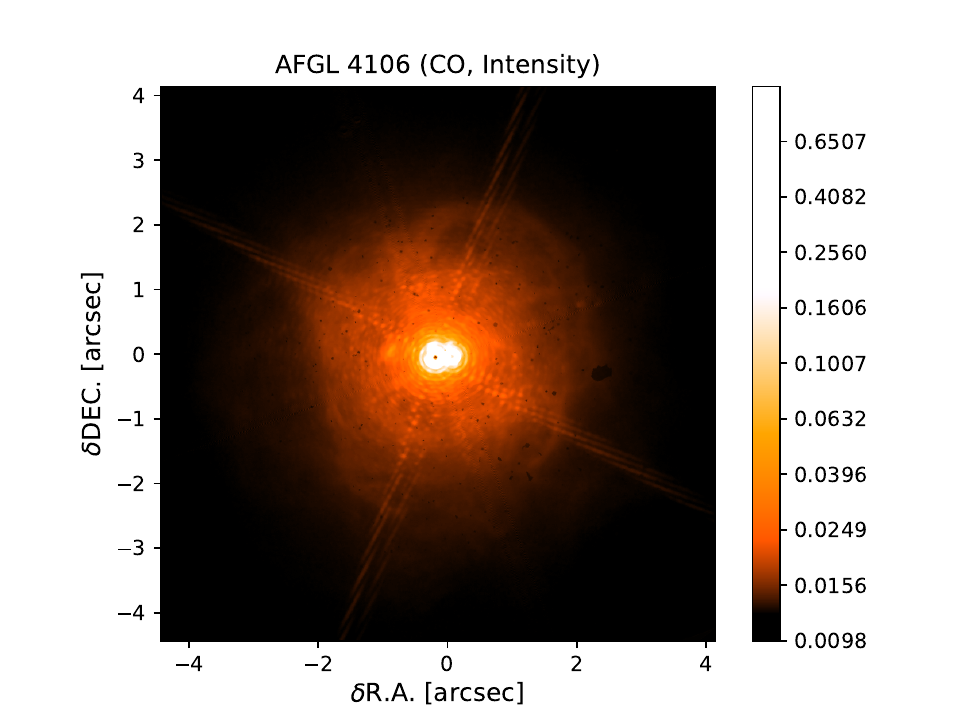}  
    \caption{IRDIS intensity map of AFGL 4106 in filter CO.}
    \label{fig:CO}
\end{figure}

\newpage
\onecolumn

\end{document}